\documentclass[aps,longbibliography,pre,twocolumn,bibnotes,superscriptaddress,floatfix]{revtex4-1}

\usepackage{amsmath}
\usepackage{mathtools}
\usepackage{amsthm}
\usepackage{amsfonts}
\usepackage{amssymb}
\usepackage{dcolumn}
\usepackage{bm, amssymb, color}
\usepackage[colorlinks=true,linkcolor=blue,citecolor=blue,urlcolor=black,bookmarks,breaklinks=true]{hyperref}
\usepackage{natbib}
\usepackage{tabulary}
\usepackage{multirow}
\usepackage{amsmath,soul}
\usepackage[export]{adjustbox}
\usepackage{float}
\usepackage[caption=false]{subfig}
\usepackage{array}
\usepackage{cases}
\usepackage{makecell}
\usepackage{physics}
\usepackage{yhmath}
\usepackage{enumitem}
\usepackage{soul}
\usepackage{placeins}
\usepackage{comment}

\usepackage[colorinlistoftodos,shadow,textwidth=18mm]{todonotes}

\usepackage{soul,xcolor}
\setstcolor{red}

\usepackage[colorinlistoftodos,shadow,textwidth=18mm]{todonotes}

\begin{document}


\title{Precise Determination of  Pair Interactions  from Pair Statistics of Many-Body Systems In and Out of Equilibrium}

\author{Salvatore Torquato}
\email[]{Email: torquato@princeton.edu}
\affiliation{Department of Chemistry, Department of Physics, Princeton Institute of Materials, and Program in Applied and Computational Mathematics Princeton University, Princeton, New Jersey 08544, USA}
\affiliation{School of Natural Sciences, Institute for Advanced Study, 1 Einstein Drive, Princeton, NJ 08540, USA}

\author{Haina Wang}
\affiliation{Department of Chemistry, Princeton University, Princeton, NJ 08544, USA}

\date{\today}

\begin{abstract}

The determination of the pair potential $v({\bf r})$ that accurately yields an equilibrium state at
positive temperature $T$ with a prescribed pair correlation function $g_2({\bf r})$ or corresponding structure factor $S({\bf k})$ in $d$-dimensional Euclidean space $\mathbb{R}^d$ is an outstanding inverse statistical mechanics problem with far-reaching implications.
Recently, Zhang and Torquato  conjectured that  any realizable  $g_2({\bf r})$ or $S({\bf k})$ corresponding to a translationally invariant nonequilibrium system can be attained by a classical equilibrium ensemble involving only (up to) effective pair interactions.
Testing this conjecture for nonequilibrium systems as well as for nontrivial equilibrium states requires improved inverse methodologies. 
We have devised a novel optimization algorithm  to find effective pair potentials that correspond to pair statistics of general translationally invariant disordered many-body equilibrium or nonequilibrium systems  at positive temperatures.
This methodology utilizes a parameterized family of pointwise basis functions for the potential function whose initial form is informed by small- and large-distance behaviors dictated by statistical-mechanical theory. 
Subsequently, a nonlinear optimization technique is utilized to minimize an objective function
that incorporates both the target pair correlation function $g_2({\bf r})$ and structure factor
 $S({\bf k})$ so that the small- and large-distance correlations are very accurately captured.
To illustrate the versatility and power of our methodology, we accurately determine the effective pair interactions of the following four diverse target systems: (1) Lennard-Jones system in the vicinity of its critical point;  (2) liquid under the Dzugutov potential;
(3) nonequilibrium random sequential addition packing; and 
(4) and a nonequilibrium hyperuniform ``cloaked" uniformly randomized lattice (URL).  
We found that the optimized pair potentials generate corresponding pair statistics
that accurately match their corresponding targets with total $L_2$-norm errors that are an order of magnitude smaller than that of previous methods. The results of our investigation lend further support to the Zhang-Torquato conjecture. Furthermore, our algorithm will enable one to probe systems
with identical pair statistics but different higher-body
statistics, which would shed light on the well-known degeneracy problem of statistical mechanics.

\end{abstract}

\maketitle

\section{Introduction}
\label{intro}

The relationship between the interactions in many-body systems and their corresponding structural and equilibrium/nonequilibrium properties continues to be a subject of great fundamental and practical interest in statistical physics, condensed-matter physics, chemistry, mathematics and materials science \cite{Va06,Re06a,To09a,Co09}.
Direct computer simulation of matter at the molecular and colloidal level has generated
a long and insightful tradition; see Refs.  \cite{Fr83,St85b,Ha86,Al87,Fr96,Di99,Na98,Wa99,Ro11,Sa13,Ber15} and references therein.
The fruitful  ``forward"  approach of statistical mechanics identifies a known substance that possesses scientific and/or technological interest, creates
a manageable approximation to the interparticle interactions that operate
in that substance, and exploits molecular dynamics or Monte Carlo
algorithms to predict nontrivial details concerning the structure, thermodynamics and kinetic
features of the system.  Inverse statistical-mechanical
methods \cite{To09a} allow  for  a  new  mode  of  thinking  about  the
structure and physical properties of condensed phases of matter and are ideally suited for materials discovery by design.
In the inverse approach, one attempts to determine a potential function
(subject to constraints) that robustly and spontaneously lead to {\it targeted} many-particle configurations, targeted correlation functions or a
targeted set of physical properties  over a wide range of conditions \cite{Bu74,To09a,Ja17,Sh20b}.
The targeting of equilibrium crystal states has been particularly successful, including the capacity to design low-coordinated crystals as  (zero-temperature) ground states \cite{Re05,Re07a,Ma13b,To09a,Zh13a,Jain13,Ja14} as well as crystals at positive temperature $T$ \cite{Li16,Ch18c}.

Finding the pair potential $v({\bf r})$ that accurately yields a single-component equilibrium disordered state (e.g., simple liquids and polymers) at 
positive $T$ with a prescribed pair correlation function $g_2({\bf r})$ or  corresponding structure 
factor $S({\bf k})$ in $d$-dimensional Euclidean space 
$\mathbb{R}^d$ is an outstanding inverse problem with far-reaching implications.
For example, such techniques enable one to devise atomic-based Hamiltonian models that are consistent with the experimental determination
of $S({\bf k})$, which is directly obtainable from scattering-intensity data.
In addition, one can use the solutions of such inverse problems to {\it design}, at will, portions of  the equilibrium
phase diagram of colloidal systems when {\it hypothesized} functional forms for the pair statistics
are realizable \cite{Uc06a,To03a,Kuna11,Zh20} by effective pair interactions. 

Moreover,  Zhang and Torquato \cite{Zh20} recently introduced a theoretical formalism that 
provides a means to draw equilibrium classical particle configurations from canonical ensembles with one- and two-body interactions that correspond to targeted functional forms for $g_2({\bf r})$ or $S({\bf k})$. This formalism enabled them to devise an efficient algorithm to construct systematically canonical-ensemble particle configurations with such targeted $S(\mathbf{k})$ at all wavenumbers, whenever realizable. However, 
their procedure does not provide the explicit forms of the underlying one- and two-body potentials. In the same study,
Zhang and Torquato \cite{Zh20} conjectured that  any realizable  $g_2({\bf r})$ or $S({\bf k})$ corresponding to a translationally invariant nonequilibrium  system can be attained by an equilibrium ensemble involving only (up to) effective pair interactions.
Successful solutions to the aforementioned inverse problem would enable one to test this remarkable conjecture and its implications if proved to be true.

While significant progress has been made in ascertaining pair interactions from pair statistics
\cite{Le85,Ly95,So96,Ja17,Zh15a,He18,Sh20b}, substantial computational challenges remain. Improved methods must be 
formulated to ensure a highly precise correspondence between the pair potential and the associated pair 
statistics for both equilibrium and nonequilibrium states of matter. In the case of equilibrium ensembles, 
Henderson's theorem states that in a classical homogeneous many-body system, the pair potential $v({\bf r})$ that gives rise to a given equilibrium
pair correlation function $g_2({\bf r})$ at fixed number density $\rho$ and temperature $T > 0$ is unique up to 
an additive constant \cite{He74}. While this is a powerful uniqueness theorem, in practice, it has been shown that 
very similar equilibrium pair statistics may correspond to distinctly different pair potentials \cite{Wa20},
and hence highly precise pair information is required to achieve an accurate pair potential function. 
Importantly, testing the Zhang-Torquato conjecture for systems out of equilibrium \cite{Zh20} requires computational precision that is beyond
currently available techniques. Furthermore, no current methods treat situations in which a one-body 
potential must also be incorporated to stabilize systems with certain long-ranged pair potentials
that are required to achieve ``incompressible" exotic hyperuniform states \cite{To03a,To18a}. Indeed, Buck \cite{Bu74} has noted that in cases where the interaction is long-ranged, previous approaches could  yield ambiguous solutions, i.e., different inversion approaches yield different potentials for the same scattering data. Thus, hyperuniform targets demand a completely new and improved inverse methodology. 

All previous iterative predictor-corrector methods \cite{Le85,Ly95,So96,He18}, which appear
to be the most accurate inverse procedures, begin with an initial discretized (binned) approximation of a trial pair potential,
but without considering one-body interactions. The trial pair  potential at each binned distance
is iteratively updated to attempt to reduce the difference between the target and trial pair statistics.
A popular method, called the iterative Boltzmann inversion (IBI) \cite{So96},
takes the initial trial potential as the potential of mean force  
and then iteratively updates the trial potential based on the difference between the potentials of mean force for the target and trial pair correlation functions. A similar fixed-point iteration approach proposed by Lyubartserv et al. \cite{Ly95} reaches comparable accuracy to that of IBI \cite{So96}. The best currently available scheme
is the iterative hypernetted chain inversion (IHNCI) scheme, which was introduced by Levesque, Weiss and Reatto \cite{Le85} and refined
in Ref. \cite{He18}. The IHNCI  updates to attempt to match both the target $g_2(\mathbf{r})$ and target $S(\mathbf{k})$
by iterating the Ornstein-Zernike equation  
using the hypernetted chain approximation. By construction, this procedure can only lead to an approximation to the desired
pair potential. Moreover, because all previous methods do not optimize a pair-statistic ``distance" functional, they are unable to detect  poor agreement between the target and trial pair statistics
that may arise as the simulation evolves, leading to increasingly inaccurate corresponding trial potentials.


 



To improve on previous methods and address the aforementioned new challenges, we have devised a novel optimization 
algorithm that is informed  by statistical-mechanical theory to find effective one- and two-body potentials yielding positive-$T$ equilibrium states
that correspond to pair statistics of general translationally invariant (statistically homogeneous)  disordered many-body systems in and out of equilibrium with unprecedented accuracy \cite{Note9}. \footnotetext[9]{Since our study focuses on pure, single-component homogeneous systems, it excludes coexisting-phase states and systems with an interface.}
Our methodology departs from previous techniques in several significant ways.
First, we pose the task as an optimization problem to minimize an objective function
that incorporates both the target pair correlation function $g_{2,T}({\bf r})$ and structure factor
 $S_{T}({\bf k})$ that would correspond to one-body and two-body (pair) potential. Second, unlike previous procedures, our methodology
uses a parameterized family of {\it pointwise basis functions} for the potential function $v({\bf r};{\bf a})$,
where $\bf a$ represents a supervector of parameters (Sec. \ref{meth}) whose initial form
is informed by the small- and large-distance behaviors of the targeted $g_{2,T}({\bf r})$  and $S_{T}({\bf k})$,
as dictated by statistical-mechanical theory. Pointwise potential functions are superior 
to previously employed binned potentials, since  they do not suffer from the 
accumulation of random errors during a simulation, resulting in more accurate pair interactions. Binned potentials that contain random noise can lead to unrealistic forces in
molecular dynamics and result in incorrect supercooled or quenched structures.
Third,  a nonlinear optimization technique is utilized to minimize the objective
function so that the very-small, intermediate-, and large-distance correlations are very accurately captured.
The reader is referred to Sec. \ref{meth} for algorithmic details of our inverse methodology.

\begin{figure*}[!ht]
\subfloat[]{
    \centering\includegraphics[width=6cm]{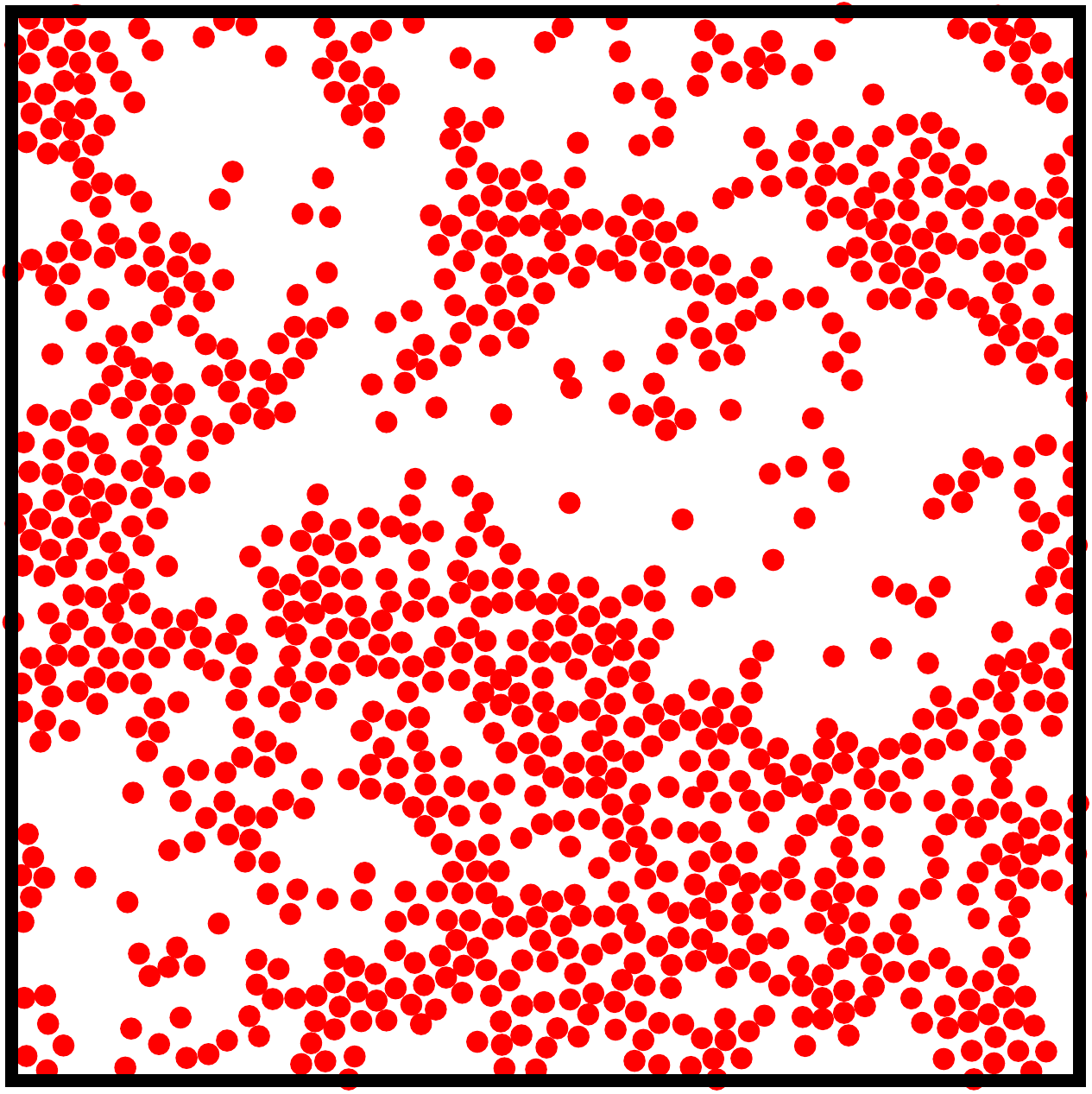}
}
\hspace{1cm}
\subfloat[]{
    \centering\includegraphics[width=6cm]{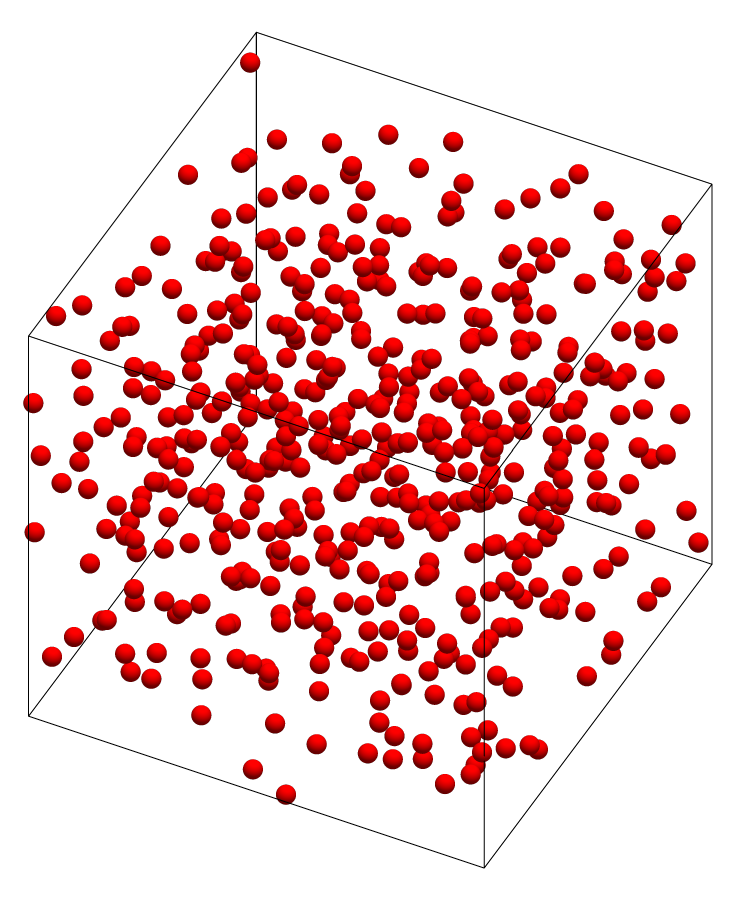}
}

\subfloat[]{
    \centering\includegraphics[width=6cm]{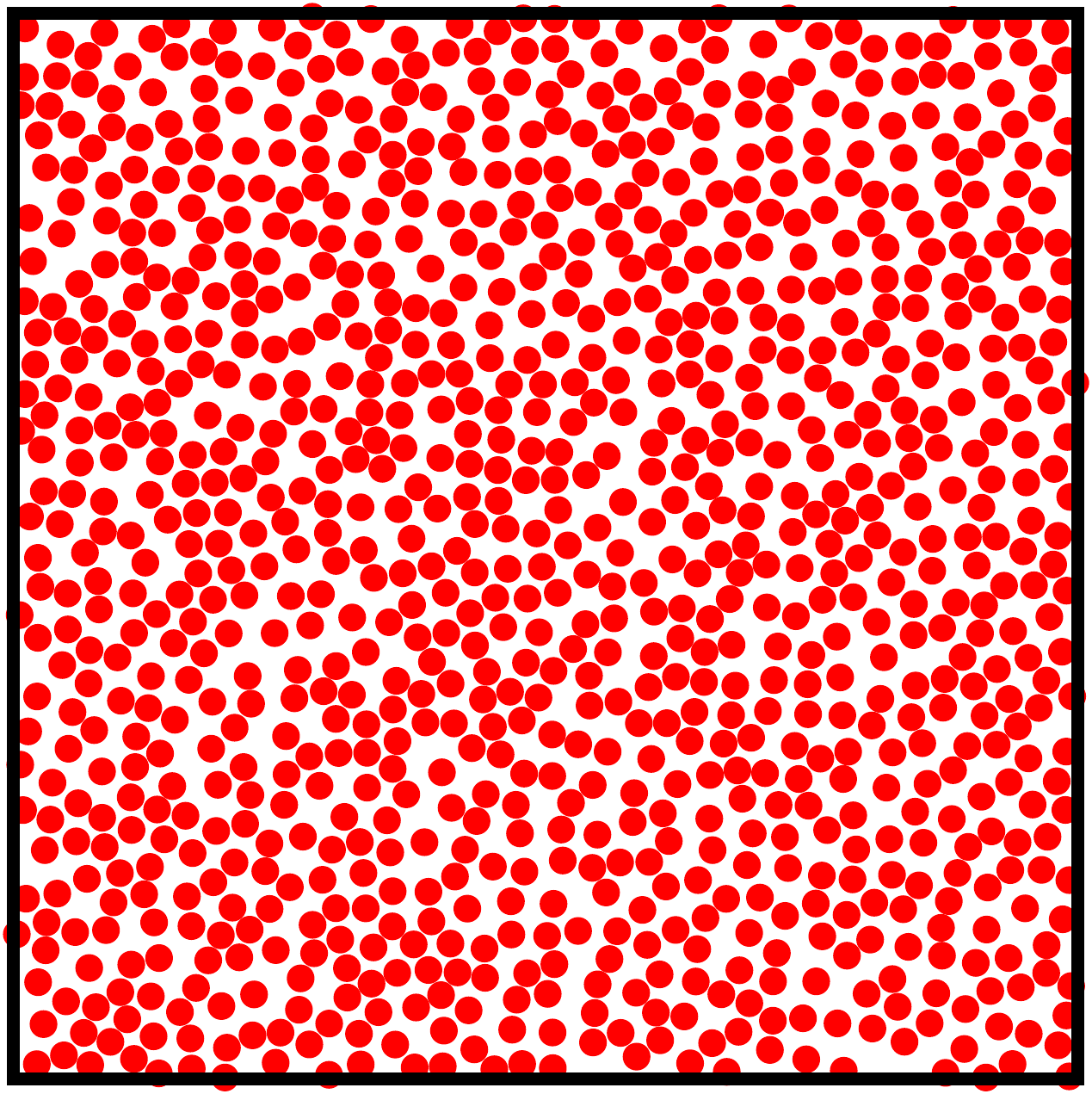}
}
\hspace{1cm}
\subfloat[]{
    \centering\includegraphics[width=6cm]{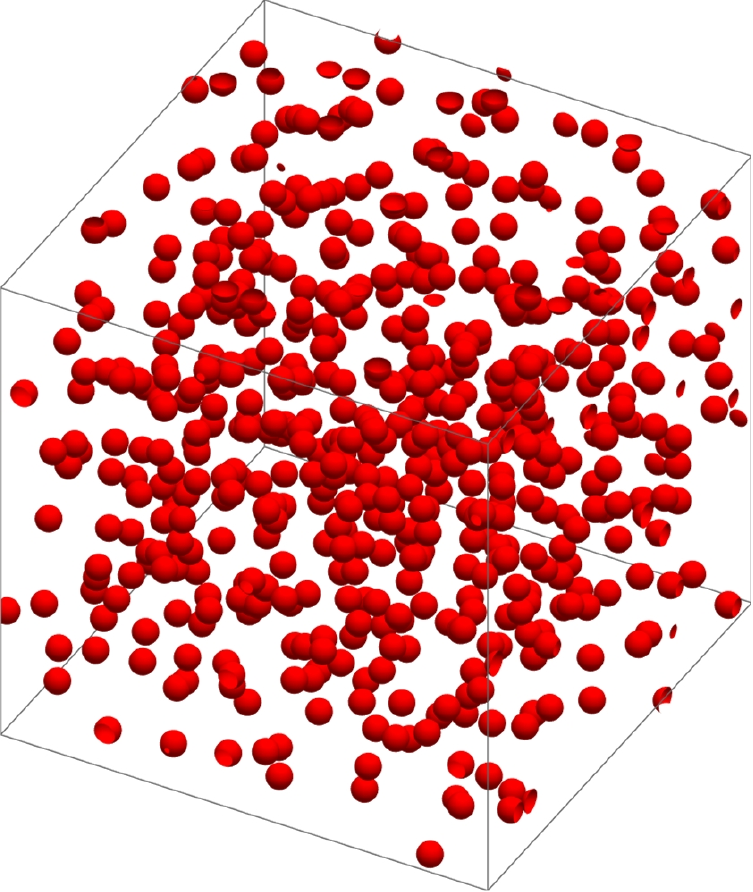}
}
\caption{Snapshots of the configurations corresponding to the four target systems considered in this study. (a) A representative 2D 1,000-particle  configuration of the target Lennard-Jones system near its critical point. (b) A representative 3D 512-particle  configuration of the target liquid under the Dzugutov potential. (c) A portion of a representative 2D 10,000-particle  RSA configuration very near the saturation state. Only 1000 particles are displayed. (d) A portion of a representative 3D 2,744-particle configuration of a cloaked URL system. Only 512 particles are displayed.}
  \label{snapTarget}
\end{figure*}

To illustrate the versatility and power of our methodology, we accurately determine the effective pair
interactions of the following four diverse target systems:
\begin{enumerate}
\item 2D Lennard-Jones system in the vicinity of its critical point;
\item 3D liquid under the Dzugutov potential \cite{Dz03};
\item 2D nonequilibrium random sequential addition (RSA) packing near saturation \cite{Fe80,Zh13b};
\item 3D nonequilibrium hyperuniform ``cloaked" uniformly randomized lattice (URL) \cite{Kl20}.
\end{enumerate}
Figure \ref{snapTarget} shows snapshots of the target configurations. These targets can be problematic for previous procedures 
to achieve accurate pair potentials for various reasons, as detailed in Sec. \ref{appli}. 

To assess the accuracy of inverse methodologies to target pair statistics, we introduce and utilize
the following dimensionless $L_2$-norm error:
\begin{equation}
{\cal E}= \sqrt{D_{g_2}+D_{S}}
\label{L2}
\end{equation}
where 
\begin{equation}
D_{g_2}=\rho\int_{\mathbb{R}^d} [g_{2,T}(\mathbf{r})-g_{2,F}(\mathbf{r};\mathbf{a})]^2 d\mathbf{r}, 
\label{g2-norm}
\end{equation}
\begin{equation}
D_S=\frac{1}{\rho (2\pi)^d}\int_{\mathbb{R}^d} [S_{T}(\mathbf{k})-S_{F}(\mathbf{k};\mathbf{a})]^2 d\mathbf{k},
\label{S-norm}
\end{equation}
where $g_{2,F}(\mathbf{r};\mathbf{a})$ and  $S_{F}(\mathbf{k};\mathbf{a})$ represent the final
pair statistics at the end of the optimization, which depend on the supervector $\bf a$.
It is noteworthy that our algorithm generally provides errors in the final pair statistics for all four targets that are an order of magnitude smaller than previous methods, as we will subsequently show. Importantly, it reaches the precision required to recover the unique potential dictated by Henderson's theorem \cite{He74}. Realizing the nonequilibrium pair statistics of RSA and URL systems by corresponding equilibrium systems with effective pair interactions lend further support to the Zhang-Torquato conjecture 
 \cite{Zh20} and shed light on the well-known degeneracy problem \cite{Ji10a,St19}. Structural, thermodynamic and dynamic properties of the equilibrium systems could also reveal nontrivial information on the nonequilibrium states, such as the degree to which the latter systems are out of equilibrium, as detailed in Sec. \ref{conclusions}.



 



 

We begin by providing basic definitions and background in Sec. \ref{def}.  In Sec. \ref{large-r}, we describe
the  asymptotic behavior of the pair potential from large-scale pair correlations for nonhyperuniform and hyperuniform targets, which are dictated by statistical-mechanical theory \cite{To18a}. Here, we also derive a general expression for the small-$|\mathbf{k}|$ behavior of $S(\mathbf{k})$ for systems equilibrated under inverse power-law pair potentials that applies in any space dimension; see Eq. (\ref{s_smallk}). In Sec. \ref{meth},  we provide a detailed
description of  our inverse methodology that is generally applicable to statistically  homogeneous but anisotropic
disordered states. In Sec. \ref{appli},  we apply  our methodology on diverse target translationally invariant
disordered systems, in and out of equilibrium.  We provide concluding remarks in Sec. \ref{conclusions}.

\section{Basic Definitions and Background}
\label{def}

\subsection{Pair Statistics}

We consider many-particle systems in $d$-dimensional Euclidean space $\mathbb{R}^d$ that are completely statistically characterized by the $n$-particle probability density functions $\rho_n(\mathbf{r}_1,...,\mathbf{r}_n)$ for all $n\geq 1$ \cite{Ha86}. Our primary interest
in this paper is in the one- and two-body statistics. In the case of statistically homogeneous systems, $\rho_1(\mathbf{r}_1)=\rho$ and $\rho_2(\mathbf{r}_1,\mathbf{r}_2)=\rho^2 g_2(\mathbf{r})$, $\rho$ is the number
density in the thermodynamic limit, $g_2(\mathbf{r})$ is the pair correlation function, and $\mathbf{r}=\mathbf{r}_2-\mathbf{r}_1$.
If the system is also statistically isotropic, then $g_2(\mathbf{r})$ is the radial function $g_2(r)$, where $r=|\mathbf{r}|$. The ensemble-averaged structure factor $S(\mathbf{k})$ is defined as
	\begin{equation}
		S(\mathbf{k})=1+\rho \tilde{h}(\mathbf{k}),
		\label{skdef}
	\end{equation}
where $h(\mathbf{r})=g_2(\mathbf{r})-1$ is the total correlation function, and $\tilde{h}(\mathbf{k})$ is the Fourier transform of $h(\mathbf{r})$.
The Fourier transform of a function $f(\mathbf{r})$ that depends on the vector $\bf r$ in $\mathbb{R}^d$ is give by
\begin{equation}
    \tilde{f}(\mathbf{k})=\int_{\mathbb{R}^d}f(\mathbf{r})\exp[-i\mathbf{k}\cdot\mathbf{r}]d\mathbf{r},
\end{equation}
where $\mathbf{k}\cdot \mathbf{r}=\sum_{i=1}^d k_ir_i$ is the conventional Euclidean inner product of two real-valued vectors. When it is well-defined, the corresponding inverse
Fourier transform is given by
\begin{equation}
    f(\mathbf{r})=\left(\frac{1}{2\pi}\right)^d\int_{\mathbb{R}^d}\tilde{f}(\mathbf{k})\exp[i\mathbf{k}\cdot\mathbf{r}]d\mathbf{k}.
\end{equation}
If $f$ is a radial function, i.e., $f$ depends only on the modulus $r = |\mathbf{r}|$ of the vector $\mathbf{r}$, its Fourier transform is given by
\begin{equation}
    \tilde{f}(k)=(2\pi)^{\frac{d}{2}}\int_0^\infty r^{d-1} f(r)\frac{J_{(d/2)-1}(kr)}{(kr)^{(d/2)-1}}dr,
\end{equation}
where $J_\nu(x)$ is the Bessel function of the first kind of order $\nu$. The inverse Fourier transform of $\tilde{f}(k)$ is given by
\begin{equation}
    f(r)=\left(\frac{1}{2\pi}\right)^{\frac{d}{2}}\int_0^\infty k^{d-1}\tilde{f}(k)\frac{J_{(d/2)-1}(kr)}{(kr)^{(d/2)-1}}dk.
    \label{inverse_fourier_radial}
\end{equation}

For a single periodic configuration containing number $N$ point particles at positions ${\bf r}_1,{\bf r}_2,\ldots,{\bf r}_N$ within a fundamental cell $F$ of a lattice $\Lambda$, the {\it scattering intensity} $\mathcal{I}(\mathbf{k})$  is defined as
	\begin{equation}
		\mathcal{I}(\mathbf{k})=\frac{\left|\sum_{i=1}^{N}e^{-i\mathbf{k}\cdot\mathbf{r}_i}\right|^2}{N}.
		\label{scattering}
	\end{equation}
For an ensemble of periodic configurations of $N$ particles within the fundamental cell $F$, the ensemble average of the scattering intensity in the infinite-volume limit is directly related to structure factor $S(\mathbf{k})$ by
	\begin{equation}
		\lim_{N,V_F\rightarrow\infty}\langle\mathcal{I}(\mathbf{k})\rangle=(2\pi)^d\rho\delta(\mathbf{k})+S(\mathbf{k}),
	\end{equation}
where $V_F$ is the volume of the fundamental cell and $\delta$ is the Dirac delta function \cite{To18a}. In simulations of many-body systems with finite $N$ under periodic boundary conditions, Eq. (\ref{scattering}) is used to compute $S(\mathbf{k})$ directly by averaging over configurations.

\subsection{Hyperuniformity and Nonhyperuniformity}

A \textit{hyperuniform} point configuration in $d$-dimensional Euclidean space $\mathbb{R}^d$ is characterized by an anomalous suppression of large-scale density fluctuations relative to those in typical disordered systems, such as liquids and structural glasses \cite{To03a,To18a}. More precisely, a hyperuniform point pattern is one in which the structure factor $S(\mathbf{k}) = 1 + \rho\tilde{h}(\mathbf{k})$ tends to zero as the wave number $k = |\mathbf{k}|$ tends to zero \cite{To03a,To18a} i.e.,
\begin{equation}
    \lim_{|\mathbf{k}|\rightarrow 0} S(\mathbf{k}) = 0.
    \label{hyperuniformDef}
\end{equation}
This hyperuniformity condition implies the following direct-space sum rule:
\begin{equation}
    \rho \int_{\mathbb{R}^d}h(\mathbf{r})d\mathbf{r} = -1.
\end{equation}
An equivalent definition of hyperuniformity is based on the local number variance $\sigma^2(r)=\langle N(R)^2\rangle - \langle N(R)\rangle^2$ associated with the number $N(R)$ of points within a $d$-dimensional spherical observation window of radius $R$, where angular brackets denote an ensemble average. A point pattern in $\mathbb{R}^d$ is hyperuniform if its variance grows in the large-$R$ limit slower than $R^d$ \cite{To03a}. 

Consider systems that are characterized by a structure factor with a radial power-law form in the vicinity of the origin, i.e.,
\begin{equation}
    S(\mathbf{k})\sim |\mathbf{k}|^\alpha \qquad \mbox{for } |\mathbf{k}| \rightarrow 0.
    \label{s_hu_smallk}
\end{equation}
For hyperuniform systems, the exponent $\alpha$ is positive ($\alpha > 0$) and its value determines three different large-$R$ scaling behaviors of the number variance \cite{To03a,Za09,To18a}: 
\begin{equation}
    \sigma^2(R)\sim \begin{cases}
          R^{d-1} \quad \alpha > 1 \text{ (class I)} \\
          R^{d-1}\ln R \quad \alpha=1 \text{ (class II)} \\
          R^{d-\alpha} \quad 0<\alpha<1 \text{ (class III).} \\
     \end{cases}
\end{equation}

By contrast, for any \textit{nonhyperuniform} system, the local variance has the following large-$R$ scaling behaviors \cite{To21c}:
\begin{equation}
    \sigma^2(R)\sim \begin{cases}
          R^{d} \quad \alpha = 0 \text{ (typical nonhyperuniform)} \\
          R^{d-\alpha} \quad \alpha<0 \text{ (antihyperuniform).} \\
     \end{cases}
\end{equation}
For a ``typical” nonhyperuniform system, $S(0)$ is bounded. In antihyperuniform systems \cite{To18a}, $S(0)$ is unbounded, i.e.,
\begin{equation}
    \lim_{|\mathbf{k}|\rightarrow 0} S(\mathbf{k}) = \infty,
\end{equation}
and hence are diametrically opposite to hyperuniform systems. Antihyperuniform systems include fractals, systems at thermal critical points (e.g., liquid-vapor and magnetic critical points) \cite{Wi65,Ka66,Fi67,Wi74,Bi92}, as well as certain substitution tilings \cite{Og19}.
A \textit{hyposurficial} state  is a special nonhyperuniform system that lacks a ``surface-area'' term in the growth of the number variance and have pair statistics that obeys the following sum rule \cite{To03a}:
\begin{equation}
    \int_0^\infty r^d h(r) dr = 0,
    \label{hypodef}
\end{equation}
which implies that they must generally contain both negative and positive correlations.
In equilibrium systems, hyposurficial states arise in the supercritical region of the phase
diagram. Hyposurficial states have been also shown to arise in computer simulations of phase transitions involving amorphous ices.\cite{Mar17}

\subsection{Hyperuniform Equilibrium States}
\label{hyperuniformEquilStates}
Torquato \cite{To18a} utilized the well-known fluctuation-compressibility relation for single-component 
equilibrium systems \cite{Ha86}
\begin{equation}
\rho k_B T \kappa_T= S(0),
\label{comp}
\end{equation} 
to infer salient conclusions about equilibrium hyperuniform states at $T=0$ (ground states)
and at positive temperatures. Specifically, any ground state ($T = 0$), ordered or disordered, in which the isothermal compressibility 
$\kappa_T$ is bounded and positive must be hyperuniform. This is true more generally if   the product $T\kappa_T$ tends to a nonnegative constant
in the limit $T \to 0$. Moreover, in order to have a hyperuniform system that is 
in equilibrium at any positive $T$,  $\kappa_T$ must be identically zero, i.e., the system must be thermodynamically incompressible, at least
classically.
Such a situation requires an effective pair potential that is unusually long-ranged, as will be specifically shown 
in Sec. \ref{large-r}.

\section{Asymptotic Behavior of the Pair Potential from Large-Scale Pair Correlations}
\label{large-r}
In this section, we describe how to infer the large-distance behavior of the pair potential from targeted large-scale pair correlations for nonhyperuniform and hyperuniform states. We also derive a new expression (\ref{s_smallk}) for the small-$|\mathbf{k}|$ behavior of $S(\mathbf{k})$ for systems equilibrated under inverse power-law pair potentials in any space dimension.

For single-component equilibrium fluids, the asymptotic behavior of the direct correlation function $c({\bf r})$, defined via the Ornstein-Zernike integral equation \cite{Ha86},
for large $|\bf r|$ determines the  asymptotic behavior of the pair potential $v({\bf r})$, provided that $h^2({\bf r}) \ll |\beta v({\bf r})|$ \cite{St77}; specifically, 
we have
\begin{equation}
c({\bf r}) \sim -\beta v({\bf r}) \quad (|{\bf r}| \to \infty),
\label{asymp}
\end{equation}
where $\beta =1/(k_B T)$ and $k_B$ is the Boltzmann constant. 
While the condition  $h^2({\bf r}) \ll |\beta v({\bf r})|$ is violated at a liquid-vapor critical point (since $h({\bf r}) \sim 1/|{\bf r}|^{d-2+\eta}$ in the large-$|\bf r|$ limit,
where $\eta=1/4$ for $d=2$ and $\eta\approx 0.036$ for $d=3$ \cite{Bi92}), it is not violated
away from and near the critical point. The latter conclusion follows from the well-known Ornstein-Zernike
analysis that in the vicinity of the critical point, the structure factor takes the following Lorentzian form
in the infinite-wavelength limit \cite{Fi64b, Fi67}:
\begin{equation}
    S({\bf k})=\frac{S(0)}{1 + \xi^2 |{\bf k}|^2 + O(k^4)}\quad |{\bf k}|\rightarrow 0,
\label{lorentz}
\end{equation}
where $\xi$ is the correlation length. It then immediately follows that the total correlation function
has the large-$r$ form
\begin{equation}
    h({\bf r})\sim \frac{\exp(-|{\bf r}|/\xi)}{|{\bf r}|^{(d-1)/2}} \qquad (|{\bf r}|\rightarrow \infty).
    \label{crit_h}
\end{equation}
At the critical density $\rho=\rho_c$, $\xi$ tends to infinity as the temperature approaches the critical temperature $T_c$ via the scaling relation \cite{Fi67}
\begin{equation}
    \xi\propto\left(\frac{T-T_c}{T_c}\right)^{-\nu} \qquad T\rightarrow T_c^+,
\label{nu}
\end{equation}
where $\nu=1$ for $d=2$ and $\nu\approx 0.63$ for $d=3$ \cite{Bi92}.
Thus, in light of (\ref{crit_h}) and (\ref{nu}), we can conclude 
(\ref{asymp}) applies in the vicinity of a critical point for any bounded
correlation length $\xi$.

The large-$|\bf r|$ behavior of $c({\bf r})$ or $v({\bf r})$ via (\ref{asymp}) can be extracted
from the structure factor $S({\bf k})$ 
using the Fourier representation of the Ornstein-Zernike integral equation \cite{Or14}:
\begin{equation}
{\tilde c}({\bf k})= \frac{{\tilde h}({\bf k})}{S({\bf k})}
\label{OZ}
\end{equation}
and examining the small-wavenumber behavior of ${\tilde c}({\bf k})$, which is the Fourier transform of $c({\bf r})$.
If ${\tilde c}({\bf k})$ is analytic at the origin, i.e., the Taylor series about $|{\bf k}|=0$ involves only even powers
of the wavenumber $|{\bf k}|$, it immediately follows
that $c({\bf r})$ decays exponentially fast or faster \cite{To18a} and hence 
via (\ref{asymp}), $v({\bf r})$ must decay  exponentially fast or faster, i.e., faster than any inverse power law.
However, if the pair potential $v({\bf r})$ for large $\bf r$ is asymptotically given by the following power-law form:
\begin{equation}
    v({\bf r})\sim \frac{E}{|{\bf r}|^{d+\zeta}} \qquad (|{\bf r}|\rightarrow \infty)
    \label{power_law_pot}
\end{equation}
where $E$ is a constant with units of energy that may be positive or negative, $|{\bf r}|$ is a dimensionless pair distance and $\zeta > 0$,
then according to (\ref{asymp}), we have  
\begin{equation}
    c({\bf r})\sim -\beta v({\bf r})=-\frac{\beta E}{|{\bf r}|^{d+\zeta}} \qquad (|{\bf r}|\rightarrow \infty).
\label{power-law}
\end{equation}

Given a system in which the tail of the potential has the power-law form (\ref{power_law_pot}), we can extract the long-wavelength behavior of the structure factor following the asymptotic analysis of the Ornstein-Zernike equation presented in Ref. \cite{To18a}. The specific behavior of $S({\bf k})$ depends on whether $\zeta$ is an odd integer, even integer or fractional. When $\zeta$ is an odd integer, the structure factor in the long-wavelength limit consists of nonanalytic and analytic contributions:
\begin{equation}
S({\bf k})\sim 
    -\frac{E}{k_B T}C_1(\zeta, d)\rho S(0)^2 |{\bf k}|^\zeta + f({\bf k}) 
\quad (|{\bf k}|\rightarrow 0),
\label{s_smallk}
\end{equation}
where
\begin{equation}
    C_1(\zeta, d)=\frac{\pi^{1+d/2}}{2^\zeta \Gamma(1+\zeta/2)\Gamma((d+\zeta)/2)\sin(\pi \zeta/2)}
\end{equation}
and $f({\bf k})$ is an analytic function of $|{\bf k}|$ at the origin of the form
\begin{equation}
    f({\bf k})=s_0+s_2 |{\bf k}|^2+s_4 |{\bf k}|^4 + s_6 |{\bf k}|^6 + ...
\end{equation}
Whereas $s_0 = S(0)$, independent of the value of the positive exponent $\zeta$, the coefficients $s_{2m}$ for $m\geq 1$ depend on $\zeta$ and a subset of them may be expressed as even-order moments of the total correlation function $h({\bf r})$. When $\zeta$ is an even integer, the asymptotic form (\ref{s_smallk}) no longer applies, but the asymptotic formula now includes a nonanalytic term proportional to $\ln(|{\bf k}|)|{\bf k}|^{\zeta}$.

In the case of an equilibrium hyperuniform state whose structure factor is characterized by the power law (\ref{s_hu_smallk}), the Ornstein-Zernike equation (\ref{OZ}) implies that $\tilde{c}(\mathbf{k})$ contains a singularity at the origin \cite{To18a}:
\begin{equation}
    \tilde{c}(\mathbf{k})\sim -|\mathbf{k}|^{-\alpha} \qquad |\mathbf{k}|\rightarrow 0.
    \label{ck_hu}
\end{equation}
Fourier transformation of Eq. (\ref{ck_hu}) shows that $c(\mathbf{r})$ and hence $v(r)$ have the inverse power-law decay \cite{To18a}
\begin{equation}
    c(\mathbf{r})\sim -\beta v(\mathbf{r})\sim -|\mathbf{r}|^{-(d-\alpha)} \qquad |\mathbf{r}|\rightarrow\infty.
    \label{v_hu}
\end{equation}
Thus, $v(r)$ is long-ranged in the sense that its volume integral is unbounded. 
Therefore, to stabilize a  classical hyperuniform system at positive $T$, which is thermodynamically incompressible due to Eq. (\ref{comp}), one must treat it as a system of ``like-charged'' particles immersed in a rigid ``background'' of equal and opposite ``charge'', i.e., the system must have overall charge neutrality \cite{To18a}. Such a rigid background contribution corresponds to a one-body potential, which 
is described in detail in Sec. \ref{meth_hu}.

\section{New Inverse Methodology}
\label{meth}

For present purposes, we consider the canonical ensemble associated with classical many-body systems of $N$ particles in a finite but large region $\Omega\subset \mathbb{R}^d$ of volume $V$ in thermal equilibrium with a heat bath at absolute temperature $T$. We assume that the particles are subject to a one-body and a two-body potential,  $\phi_1(\mathbf{r})$ and $\phi_2(\mathbf{r}_1,\mathbf{r}_2)$, respectively. Thus, the configurational potential energy of particles at $\mathbf{r}^N=\{\mathbf{r}_1, \mathbf{r}_2, \dots, \mathbf{r}_N\}$ is given by
\begin{equation}
    \Phi(\mathbf{r}^N)=\sum_{i=1}^N \phi_1(\mathbf{r}_i)+\frac{1}{2}\sum_{i=1}^N\sum_{j\ne i}^N \phi_2(\mathbf{r}_i, \mathbf{r}_j).
    \label{Phi_config}
\end{equation}
In order to have a system with a well-defined
thermodynamic behavior,  we consider $\phi_1$ and $\phi_2$ that are \textit{stable}, i.e., there exist positive constants $B_1, B_2\geq 0$, such that for all $N$ and all $\mathbf{r}^N\in \mathbb{R}^{Nd}$ \cite{Ru70}
\begin{equation}
    \sum_{i=1}^N \phi_1(\mathbf{r}_i) \geq -NB_1
\end{equation}
and
\begin{equation}
    \frac{1}{2}\sum_{i=1}^N\sum_{j\ne i}^N \phi_2(\mathbf{r}_i, \mathbf{r}_j)\geq -NB_2.
\end{equation}
Ultimately, we are interested in the thermodynamic limit, i.e., the number density $\rho =N/V$
is a fixed constant in the limits $N \to \infty$ and $V\to \infty$.
Due to an extension of Henderson's theorem, the one- and two-body potentials $\{\phi_1(\mathbf{r}), \phi_2(\mathbf{r}_i,\mathbf{r}_j)\}$ that give rise to a given set of equilibrium probability density functions $\{\rho_1(\mathbf{r}), \rho_2(\mathbf{r}_i, \mathbf{r}_j)\}$ at fixed positive and finite temperature $T$ are unique up to additive constants \cite{Ch84b,Wa20}.


In what follows,  we describe our new inverse methodology to find one- and two-body potentials $\phi_1(\mathbf{r})$ and $\phi_2(\mathbf{r}_1,\mathbf{r}_2)$ at some positive and finite temperature $T$ such that the corresponding equilibrium one- and two-body probability density functions accurately match the target one- and two-body statistics.  Our multi-stage procedure consists of choosing the initial forms of
pointwise basis functions for the parameterized potentials,
which is informed by statistical-mechanical theory described in Sec. \ref{large-r}, and then we iteratively optimize the potential parameters
until a convergence criterion is attained.
If we do not achieve sufficiently accurate solutions in the first stage, then we choose a new set of basis functions by refining
the previous choice and iteratively proceed until our convergence criterion is attained, as detailed below.

\subsection{Objective function}
\label{obj}
Here, we describe the objective function that must be minimized for the inverse procedure. 
We generally assume statistically anisotropic target pair statistics
that are vector-dependent, i.e.,  $g_{2,T}({\bf r})$ and $S_{T}({\bf k})$, and hence the associated pair potential
$\phi_2({\bf r})$ is directionally-dependent (anisotropic) and one-body potential $\phi_1$ is uniform (constant).  Since it has recently been established \cite{Wa20} that methods that target only 
$g_2({\bf r})$ or only $S({\bf k})$ may generate effective potentials that are distinctly different from the unique pair potential dictated by Henderson's theorem, we consider an objective function that involves a ``distance'' functional incorporating both functions. Specifically,  given  the target pair statistics $g_{2,T}({\bf r})$ and $S_{T}({\bf k})$,
we optimize the following objective function $\Psi({\bf a})$ over the ``supervector'' parameter ${\bf a}$:
\begin{widetext}
\begin{equation}
        \Psi(\mathbf{a}) =\rho\int_{\mathbb{R}^d}
 w_{g_2}({\bf r})\left(g_{2,T}({\bf r})-g_{2}({\bf r};\mathbf{a})\right)^2
d{\bf r} + \frac{1}{\rho(2\pi)^d}\int_{\mathbb{R}^d} w_{S}({\bf k})\left(S_T({\bf k})-S(k;\mathbf{a})\right)^2 d\mathbf{k},
\label{Psi}
\end{equation}
\end{widetext}
where $g_{2}({\bf r};\mathbf{a})$ and $S(\mathbf{k};\mathbf{a})$ correspond to an equilibrated $N$-particle system under $v({\bf r};\mathbf{a})$ at temperature $T$, which can be obtained from Monte-Carlo or molecular dynamics simulations under periodic boundary conditions, and $w_{g_2}({\bf r})$ and $w_{S}({\bf k})$ are weight functions. 
Here $\bf a$ is a supervector of parameters, described more precisely below. 
The weight functions are chosen so that we accurately attain the targeted small-$r$ and small-$k$ behaviors of $g_{2,T}({\bf r})$ and $S_T({\bf k})$, respectively. In this work, we use Gaussians for both $w_{g_2}$ and $w_{S}$:
\begin{equation}
    w_{g_2}({\bf r}) = \exp\left[-\left(\frac{r}{\sigma_{g_2}}\right)^2\right],
\end{equation}
\begin{equation}
    w_{S}({\bf k}) = A\exp\left[-\left(\frac{k}{\sigma_{S}}\right)^2\right],
\end{equation}
where the constants $\sigma_{g_2}$ and $\sigma_{S}$ regulate the targeted ranges of $g_{2,T}({\bf r})$ and $S_T({\bf k})$, respectively, the constant $A$ is added to ensure that both terms in Eq. (\ref{Psi}) are of the same order of magnitude, $r \equiv |\bf r|$ and $k \equiv |\bf k|$.

\subsection{Choosing basis functions for the parameterized pair potential}
\label{parapot}

In what follows, the formulation  considers target structures that are both statistically homogeneous and isotropic. Therefore, 
the one-body potential $\phi_1(\mathbf{r})$ is  independent of $\mathbf{r}$ and $\phi_2(\mathbf{r}_i,\mathbf{r}_j)$ is a radial potential function:
\begin{equation}
    v(r_{ij})=v(|\mathbf{r}_i-\mathbf{r}_j|)=\phi_2(\mathbf{r}_i,\mathbf{r}_j),
\end{equation}
where $r_{ij}=|\mathbf{r}_i-\mathbf{r}_j|$.
The generalization of the methodology to directional or anisotropic pair potentials is described in Sec. \ref{aniso}.
We begin by considering the choice of basis functions for the pair potential $v(r)$
for nonhyperuniform targets, which means the one-body term in Eq. (\ref{Phi_config}) 
is irrelevant and so here we set  $\phi_1=0$.  
Consider a parameterized isotropic potential function $v(r;{\bf a})$ that we decompose into  a sum of $n$ smooth pointwise basis functions, i.e.,
\begin{equation} 
v(r;{\bf a})= \varepsilon \sum_{j=1}^n f_j(r/D;a_j), 
\label{basis} 
\end{equation} 
where $f_j(r/\sigma;a_j)$ is the $j$th basis function, $a_j$ is a vector of parameters (generally consisting of multiple parameters),
${\bf a}=(a_1,a_2,\ldots,a_n)$ is the ``supervector'' parameter, $\varepsilon$ sets the energy scale and $D$
is a characteristic length scale, which in the ensuing discussion is taken to be unity, $D=1$.
The components of $a_j$ are of four types: dimensionless energy scales $\varepsilon_j$, dimensionless distance scales $\sigma_j$, dimensionless phases $\theta_j$, as well as dimensionless exponents $p_j$.

The basis functions are chosen so that they reasonably span all potential functions that could correspond to a targeted $g_{2,T}(r)$ for all $r$ or a targeted $S_T(k)$ for all $k$ under the constraint that
the resulting potential function $v(r)$ satisfies the small-$r$ and large-$r$ behaviors dictated by $g_{2,T}(r)$ and $S_T(k)$, as detailed below. For our specific targets, $f_j(r;a_j)$ are chosen from the possible following general forms:
\begin{enumerate}
\item Hard core:
\begin{equation}
f_j(r;a_j)=
    \begin{cases}
    \infty \qquad &r \leq \sigma_j \\
    0 \qquad &r > \sigma_j.
    \end{cases}
    \label{hard_core}
\end{equation}
\item Gamma-damped oscillatory form:
\begin{equation}
    f_j(r;a_j)=\frac{\varepsilon_{j}\cos\left(r/\sigma^{(1)}_j + \theta_j \right)}{\Gamma\left(r/\sigma^{(2)}_j\right)}.
    \label{gamma_form}
\end{equation}
\item Exponential-damped oscillatory form:
\begin{equation}
    f_j(r;a_j)=\varepsilon_j\cos\left(\frac{r}{\sigma^{(1)}_j} + \theta_j \right)\exp\left[-\left(\frac{r-\sigma^{(2)}_j}{\sigma^{(3)}_j}\right)^M\right].
    \label{exponential_form}
\end{equation}
\item Yukawa-damped oscillatory form \cite{Yu55}:
\begin{equation}
    f_j(r;a_j)=\varepsilon_j\cos\left(\frac{r}{\sigma^{(1)}_j} + \theta_j\right)\exp\left[-\left(\frac{r - \sigma^{(2)}_j}{\sigma^{(3)}_j}\right)^M\right]r^{-p_j}.
    \label{Yukawa_form}
\end{equation}
\item Power-law-damped oscillatory form:
\begin{equation}
    f_j(r;a_j)=\varepsilon_j\cos\left(\frac{r}{\sigma^{(1)}_j} + \theta_j\right)r^{-p_j},
    \label{power_law}
\end{equation}
\end{enumerate}
In this work, for simplicity and efficiency, the exponent $M$ in Eqs. (\ref{exponential_form}) and (\ref{Yukawa_form}) is restricted to be an integer that remains fixed during the optimization process described in Sec. \ref{opt}. Of course, other target structures may require the incorporation of other basis functions.

In choosing the basis functions, we ensure that the asymptotic decay rates of all $f_j(r;a_j)$'s are no slower than that of the longest-ranged basis function determined via the large-$r$ asymptotic analysis described in Sec. \ref{large-r}. In Eqs. (\ref{gamma_form})--(\ref{power_law}), we include the oscillatory factor $\cos\left(r/\sigma^{(1)}_j + \theta_j\right)$ only if the initial form for $v(r)$, which is informed by the Ornstein-Zernike equation (\ref{OZ}) for the targeted $g_{2,T}(r)$ and $S_T(k)$, clearly exhibits oscillatory behavior, e.g., if the initial function $v(r)$ passes through the horizontal axis at three or more values of $r$.  If not, we set $\sigma^{(1)}_j=\infty$ and $\theta_j=0$. 
We regard a set of basis functions to be a good fit of $v(r)$, if the root mean square of the fit residuals is smaller than 0.01. If there is more than one set of basis functions that satisfy this condition, then we select the one with the lowest Bayesian information criterion (BIC), which is a well-established criterion for model selection that rewards goodness of fit while penalizing an increased number of free parameters \cite{Sc78}. Once the forms of the basis functions $f_j$ are chosen, we then move to the procedure of optimizing the parameter ``supervector'' $\mathbf{a}$.

The exact large-$r$ behavior of $v(r;{\bf a})$ is enforced in (\ref{basis}) to be the large-$r$ behavior of the targeted  $c_T(r)$ via relation (\ref{asymp}), which is obtained from the targeted
structure factor $S_T(k)$ and the Ornstein-Zernike equation (\ref{OZ}), as described in Sec. \ref{large-r}, 
provided that the target meets the mild condition described there. This asymptotic form for  $v(r;{\bf a})$ enables us to choose one or more
of  the basis functions $f_j(r;a_j)$ to have the
same asymptotic form. For example, one can infer from $c(r)$ whether the longest-ranged basis function that obeys the exact large-$r$ behavior is attractive or repulsive, and whether its asymptotic decay is power-law, exponential or superexponential.
To determine the initial small-$r$ behavior of $v(r;{\bf a})$, we could use  highly accurate but complicated estimates of the bridge diagrams to close the Ornstein-Zernike integral equation \cite{Ha86}. However, such exquisitely high accuracy is not required
in the initial form because the entire function $v(r;{\bf a})$ is subsequently optimized.
To estimate the initial small- and intermediate-$r$ behavior in (\ref{basis}), we simply use the hypernetted-chain (HNC) approximation \cite{Ha86}, i.e., 
 \begin{equation} 
 \beta v_{\text{HNC}}(r)= h_T(r)- c_T(r)- \ln[g_{2,T}(r)].
 \label{HNC}
\end{equation}
\smallskip

\noindent{More specifically, the initial  small-$r$ as well as intermediate-$r$ behaviors of the basis functions and their corresponding initial
potential parameters $a_j$'s are obtained via a numerical fit on the binned HNC approximation.}

\begin{figure*}
    \centering
    \includegraphics[width=130mm]{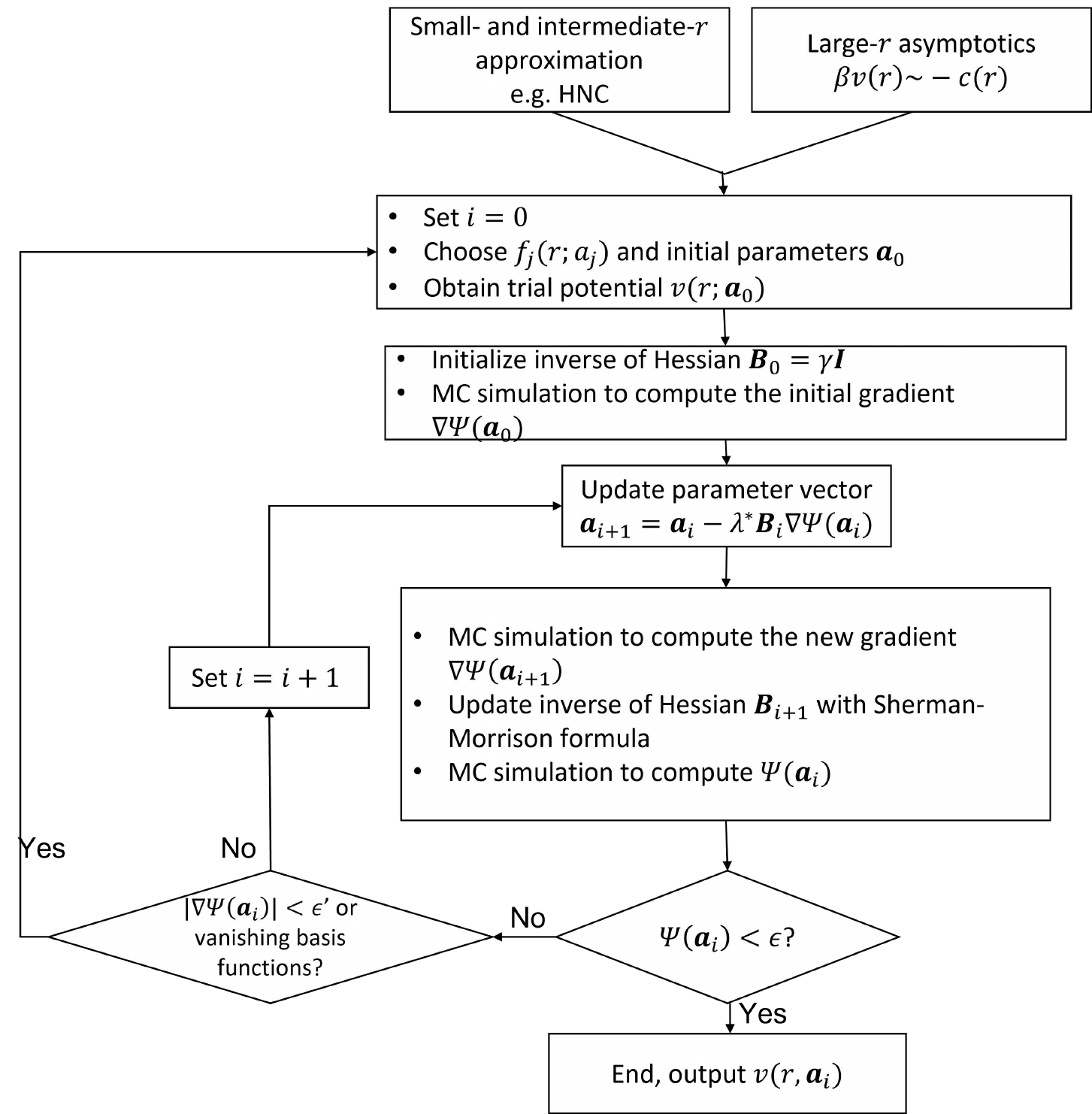}
    \caption{Flowchart of the optimization algorithm for our inverse procedure, assuming homogeneous target systems.}
    \label{fig:flowchart}
\end{figure*}

\subsection{Optimization algorithm}
\label{opt}
The second stage of our procedure involves optimizing the supervector parameter $\mathbf{a}$ in $v(r;\mathbf{a})$, such that the objective function $\Psi(\mathbf{a})$ is minimized. For this purpose, we utilize the BFGS algorithm \cite{Liu89}, which is a robust nonlinear optimization technique that we have fruitfully employed for many of our previous inverse problems \cite{Uc04b, Ba08, Zh15a}. Here we choose to equilibrate systems using Monte Carlo simulations but, of course, one may also choose to do so via molecular dynamics. Our
Our inverse algorithm consists of the following detailed steps:
\begin{enumerate}
    \item Set $i = 0$. The initial guess of the inverse Hessian matrix $\mathbf{B}_0$ is taken to be the $\gamma I$. Here, $\gamma$ is a constant which we set to be 0.1 for all systems in this study, and $I$ is the $m$ by $m$ identity matrix, where $m=|\mathbf{a}|$, i.e., the total number of \textit{scalar} components in $\mathbf{a}$.
    \item \label{compute_grad} To compute the partial derivative of $\Psi$ on the parameter $a^{(k)}$, which is the $k$-th scalar component of $\mathbf{a}$, we choose a finite difference $\delta a>0$. Monte Carlo simulations are performed under the potentials $v[r;(a^{(1)}, a^{(2)}, \cdots, a^{(k)} - \delta a/2, \cdots, a^{(m)})]$ and 
$v[r;(a^{(1)}, a^{(2)}, \cdots, a^{(k)} + \delta a/2, \cdots, a^{(m)})]$, respectively, with periodic boundary conditions, to obtain the corresponding pair statistics, namely, the standard binned ensemble-averaged $g_2(r;\mathbf{a})$ and $S(k;\mathbf{a})$ [c.f. Eq. (\ref{scattering})]
are computed.  The objective functions obtained for these two simulations are denoted $\Psi_{ik}^-$ and $\Psi_{ik}^+$, respectively. The partial derivative is computed as $(\Psi_{ik}^+ - \Psi_{ik}^-)/\delta a$. The calculations of the partial derivatives are parallelized for computational efficiency. The gradient $\nabla \Psi(\mathbf{a}_i)$ is then simply a column vector containing all the partial derivatives. In practice, it is convenient to use the same value of $\delta a$ for all parameters. We find that due to simulation errors in $\Psi$, choosing $\delta a_j$ to be too small causes large uncertainties in $\nabla \Psi (\mathbf{a}_i)$. A good choice of $\delta a_j$ is such that the uncertainty of $|\nabla \Psi (\mathbf{a}_i)|$ is less than 10\%.
    \item \label{update_a} Update the parameters as $\mathbf{a}_{i+1} = \mathbf{a}_i -\lambda^* \mathbf{B}_i  \nabla\Psi(\mathbf{a}_i)$, where $\lambda^*=\lambda\min(1, 1/|\mathbf{B}_i\nabla\Psi(\mathbf{a}_i)|)$ and $\lambda>0$ is a parameter that regulates the ``step size'' of the updates.
    \item Update the gradient $\nabla \Psi(\mathbf{a}_{i+1})$ with the same procedure as Step \ref{compute_grad}.
    \item Update the inverse Hessian $\mathbf{B}_{i+1}$ using the Sherman-Morrison formula:
\begin{eqnarray}
  \mathbf{B}_{i+1} &=&
  \mathbf{B}_i+\frac{(\mathbf{s}_i^T\mathbf{y}_i+\mathbf{y}_i^T\mathbf{B}_i\mathbf{y}_i)(\mathbf{s}_i\mathbf{s}_i^T)}{(\mathbf{s_i^T\mathbf{y}_i)}^2}
  \nonumber\\
&-&\frac{\mathbf{B}_i\mathbf{y}_i\mathbf{s}_i^T
\mathbf{s}_i\mathbf{y}_i^T\mathbf{B}_i}{\mathbf{s}_i^T\mathbf{y}_i},
\end{eqnarray}
where $\mathbf{y}_i=\nabla\Psi(\mathbf{a}_{i+1})-\nabla\Psi(\mathbf{a}_{i})$ and $\mathbf{s}_i=\mathbf{a}_{i+1}-\mathbf{a}_i$. In practice, due to the uncertainty in $\nabla\Psi(\mathbf{a}_i)$, which causes underestimation of $\mathbf{B}$, we reset the inverse Hessian to $\gamma \mathbf{I}$ if $|\mathbf{s}_i|<0.05$.
    \item Monte Carlo simulations are performed under the potential $v(r;\mathbf{a}_i)$ with periodic boundary conditions to obtain the corresponding $\Psi(\mathbf{a}_i)$. The optimization iterations stop when $\Psi(\mathbf{a}_i)$ is smaller than some tolerance $\epsilon$. 
    \item \label{change_basis} If $\Psi(\mathbf{a}_i)>\epsilon$, we check whether a different set of basis functions is needed. For example, additional basis functions are required  if the magnitude of the gradient $|\nabla \Psi(\mathbf{a}_i)|$ is smaller than some prescribed small value $\epsilon'$, because this implies that the BFGS optimization has found a minimum for $\Psi$, but this minimum does not satisfy the convergence criterion. For simplicity, we choose the additional basis functions $\{f_{n+1}(r;a_{n+1}),...,f_{p}(r;a_{p})\}$ by a numerical fit of the difference between the potentials of mean force for the target and trial pair statistics, i.e.,
    \begin{equation}
        \beta \sum_{j=n+1}^{p} f_j(r;a_j)\approx\ln\frac{g_2(r;\mathbf{a}_i)}{g_{2,T}(r)},
        \label{pmf}
    \end{equation}
    where $p>n$. The forms of the additional basis functions are again chosen from the general forms (\ref{hard_core})--(\ref{power_law}), and the fitting procedure is exactly analogous to that described in Sec. \ref{parapot}. On the other hand, if the energy-scale parameters $\varepsilon_j$ in some basis functions fall below a certain small value, which we set to be $10^{-3}$, then one can remove this basis function for efficiency. In both cases, we choose a new set of basis functions and the corresponding parameters according to Sec. \ref{parapot}. If no modification of the basis functions is needed, we set $i=i+1$, and repeat steps \ref{update_a}--\ref{change_basis}.
\end{enumerate}
We note that the parameters $\sigma_{g_2}, \sigma_{S}, A, \lambda, \delta a, \epsilon$ and $\epsilon'$ are fixed throughout the optimization process.

\subsection{Inverse procedure for hyperuniform targets}
\label{meth_hu}
Here, we describe the modification of the inverse procedure presented in Sec. \ref{parapot}--\ref{opt} to  target homogeneous systems 
that correspond to hyperuniform pair statistics in which $S(k) \sim k^{\alpha}$ in the limit $k \to 0$, where
$\alpha>0$. For such targets, one generally requires a long-ranged pair potential with the inverse-power-law form $v(r)\sim 1/r^{d-\alpha}$ for large $r$. Because  the volume integral of $v(r)$ diverges in the thermodynamic limit, a neutralizing one-body potential must be added to maintain stability \cite{To18a}. Therefore, we utilize the general total potential energy (\ref{Phi_config})
but take the one-body potential $\phi_1({\bf r})$ to be uniform with an isotropic pair potential $v(r)$.  The total configurational energy (\ref{Phi_config}) for $N$ particles in a large but finite region $\Omega$ is thus given by 
\begin{equation}
    \Phi(\mathbf{r}^N; \mathbf{a})=\sum_{i=1}^N \phi_1(\mathbf{a})+\sum_{i<j}^N v(r_{ij};\mathbf{a}),
    \label{homog_Phi_config}
\end{equation}
where $v(r;\mathbf{a})$ is still a sum of basis functions (\ref{basis}) in which the longest-ranged basis function is given by $f_{j}=a_{j}/r^{d-\alpha}$. We note that while  the formalism proposed by Zhang and Torquato \cite{Zh20}
for the realizability of pair statistics of hyperuniform targets implicitly includes a  
one-body potential, it was not explicitly determined there.

It is noteworthy that particles interacting under the long-ranged
pair potential $v(r; \mathbf{a})$ for a hyperuniform target can be regarded as a generalized  Coulombic interaction
of ``like-charged'' particles that are stabilized by  a uniform background of equal and opposite charge \cite{To18a}.
Thus, to  maintain stability, the one-body potential in a region $\Omega$ is taken to be 
\begin{equation}
    \phi_1(\mathbf{a})=-\frac{\rho}{2}\int_{\Omega} v(x;\mathbf{a})d\mathbf{x},
    \label{hu_phi1}
\end{equation}
where $d\mathbf{x}$ is a volume element in $\Omega$ and $x=|\mathbf{x}|$. Note that $\phi_1(\mathbf{a})$ is a function 
that depends on the region $\Omega$.
This use of such a background one-body term as a means to provide overall charge neutrality when the two-body potential is long-ranged 
had been employed to study numerically the one-component plasma \cite{Ha73, Ga79} as well as the Dyson log gas \cite{Dy62a}. 
One can combine the one- and two-body potentials to rewrite Eq. (\ref{homog_Phi_config}) as
\begin{equation}
\begin{split}
    \Phi(\mathbf{r}^N; \mathbf{a})&=\sum_{i<j}^N \left( v(r_{ij};\mathbf{a}) -\frac{1}{V} \int_{\Omega} v(x;\mathbf{a})d\mathbf{x}\right)\\
    &=\sum_{i<j}^N v_e(r_{ij};\mathbf{a}).
\end{split}
\label{Phi_scr}
\end{equation}
The summand, which we denote by $v_{e}(\mathbf{r};
\mathbf{a})$, can be regarded as an effective pair potential which is $v(r;\mathbf{a})$ ``screened'' by the background.

In order to obtain accurate pair statistics for such long-ranged $v(r;\mathbf{a})$ in the thermodynamic limit, one must simulate the total configurational energy corresponding to an infinitely large system in $\mathbb{R}^d$ that does not impose a cutoff on $v(r;\mathbf{a})$. For this purpose, we set $\Omega$ to be a hypercubic simulation box of side length $L$ under periodic boundary conditions
and consider its infinite periodic images. To compute the total configurational potential energy for this infinite system, we note that the effective interaction between particle $i$ and all images of particle $j$ is given by
\begin{equation}
    v_{e,\text{PBC}}(\mathbf{r}_{ij};\mathbf{a})=\sum_{\mathbf{n}}v(\mathbf{r}_{ij}+\mathbf{n}L;\mathbf{a})
    -\frac{1}{L^d}\int_{\mathbb{R}^d}v(\mathbf{r}_{ij};\mathbf{a})d\mathbf{x}
    \label{hu_vpbc}
\end{equation}
where $\mathbf{n}$ are sites of the hypercubic lattice $\mathbb{Z}^d$. In going from (\ref{Phi_scr}) to (\ref{hu_vpbc}), we have used the fact that $V=L^d$ and that the integration over the volume elements in all images of the region $\Omega$ is equivalent to a volume integral over $\mathbb{R}^d$. Note that due to the summation over lattice sites in Eq. (\ref{hu_vpbc}), $v_{e,\text{PBC}}(\mathbf{r}_{ij};\mathbf{a})$ is vector-dependent (anisotropic), even if $v_e(r_{ij};\mathbf{a})$ is isotropic.

Thus, to find one- and two-body potentials corresponding to hyperuniform target pair statistics, the inverse procedure follows exactly the same steps as those described in Sec. \ref{parapot}--\ref{opt}, except that in the optimization stage (Sec. \ref{opt}), the Monte Carlo simulations in steps 2 and 6 are performed under $v_{e,\text{PBC}}(\mathbf{r}_{ij},\mathbf{a})$, instead of $v(r;\mathbf{a})$. In practice, Eq. (\ref{hu_vpbc}) can be converted to an absolutely convergent integral and efficiently evaluated using the Ewald summation technique \cite{Ew21,Ha73}. The specific formula for the absolutely convergent integral depends on $d$ and $\alpha$ and the formula in the case $d=3,\alpha=2$ will be given in Sec. \ref{url}, where we apply our inverse procedure on the 3D cloaked URL.

\subsection{Extension of the methodology to homogeneous anisotropic systems}
\label{aniso}

Our methodology naturally extends to cases where the target system is anisotropic, e.g. nematic liquid crystals. The anisotropy of the system is reflected in the directional dependence of the target pair statistics $g_{2,T}(\mathbf{r})$ and $S_T(\mathbf{k})$. One could compute the anisotropic direct correlation function $c_T(\mathbf{r})$ via the Ornstein-Zernike equation (\ref{OZ}) that specifies the large-$|\mathbf{r}|$ behavior of the $v(\mathbf{r};\mathbf{a})$ via $\beta v(\mathbf{r})\sim -c(\mathbf{r})$ along the relevant directions, which are a set of discretized orientations of $\mathbf{r}$ that are not related by symmetry operations. Note that oftentimes, $c(\mathbf{r})$ has orientational symmetries, which reduces the number of relevant directions. Similarly, an initial guess of the small- and intermediate-$|\mathbf{r}|$ behaviors of $v(\mathbf{r};\mathbf{a})$ can be obtained via a numeric fit of the HNC approximation (\ref{HNC}).

We note that when choosing the basis functions, one should ensure that all basis functions have the same symmetry as $c(\mathbf{r})$. For example, one could choose the basis functions to be a product of a radial function and an angular function:
\begin{equation}
    f_j(\mathbf{r};a_j)=X_j(r;a_{j,X})\Theta_j(\mathbf{u};a_{j,\Theta}),
\end{equation}
where $a_{j, X}$ and $a_{j,\Theta}$ are parameter vectors for the radial function and the angular function, respectively, and $a_j=\{a_{j,X}, a_{j,\Theta}\}$. The orientation vector $\mathbf{u}$ is the unit vector in the direction of $\mathbf{r}$. The radial function $X_j(r;\mathbf{a}_{j,X})$ is selected from the general forms (\ref{hard_core})--(\ref{power_law}) and the angular function $\Theta_j(\mathbf{u};a_{j,\Theta})$ has the orientational symmetry of $c(\mathbf{r})$. For example, one can choose $\Theta_j(\mathbf{u}; a_{j,\Theta})$ to be Chebyshev polynomials
and spherical harmonics for $d=2$ and $d=3$, respectively.
The same multi-stage optimization procedure described in Sec. \ref{opt} is then applied to minimize $\Psi(\mathbf{r};\mathbf{a})$, except that the Monte Carlo simulations in steps 2 and 6 now involve anisotropic $g_2(\mathbf{r};\mathbf{a})$ and $S(\mathbf{k};\mathbf{a})$.

\section{Applications of the inverse methodology to a diverse set of target systems}
\label{appli}

In this section, we present the results of the application of our new inverse methodology for various target systems in two and three dimensions, including a 3D Dzugutov liquid \cite{Dz03}, a 2D Lennard-Jones fluid in the vicinity of the critical point, as well as pair statistics corresponding to 2D nonequilibrium RSA packing process \cite{Wi66, Fe80, To06a} and a hyperuniform 3D nonequilibrium cloaked URL \cite{Kl20}. For such diverse target pair statistics, we show that our methodology yields lower values of the $L_2$-norm error $\cal E$,
defined by (\ref{L2}), than those derived from currently available inverse procedures, including the IBI and IHCNI. 

To numerically determine target and trial pair statistics, we performed Monte Carlo simulations to generate ensembles of configurations equilibrated under the corresponding potentials. Our simulations used square or cubic boxes under periodic boundary conditions with $N = 500$ particles. We averaged simulated pair statistics for 1000 configurations to obtain $g_2(r;\mathbf{a})$ and $S(k;\mathbf{a})$. In all cases, we chose the convergence criterion for $\Psi$ to be $\epsilon= 0.002$. We include in Appendix \ref{details} the implementation details of the methodology in each case, including the specific procedure of choosing the basis functions, as well as the values chosen for the parameters $\sigma_{g_2}, \sigma_{S}, A, \lambda, \delta a$ and $\epsilon'$ in the optimization algorithm. 

\subsection{Lennard-Jones fluid in the vicinity of the critical point}
\label{crit}

\begin{figure*}[!ht]
\subfloat[]{
    \centering\includegraphics[width=5.5cm]{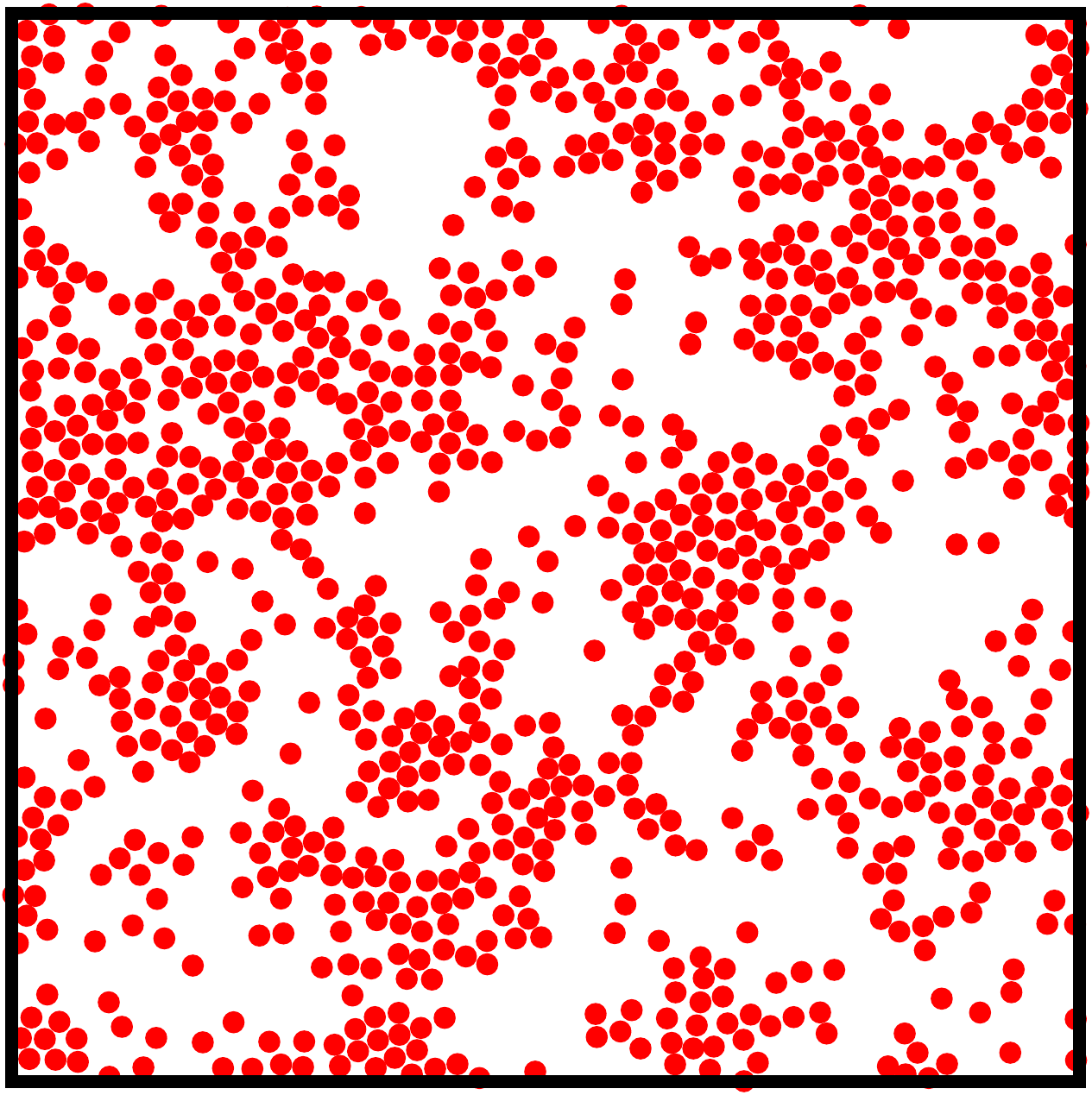}
}
\subfloat[]{
    \centering\includegraphics[width=6cm]{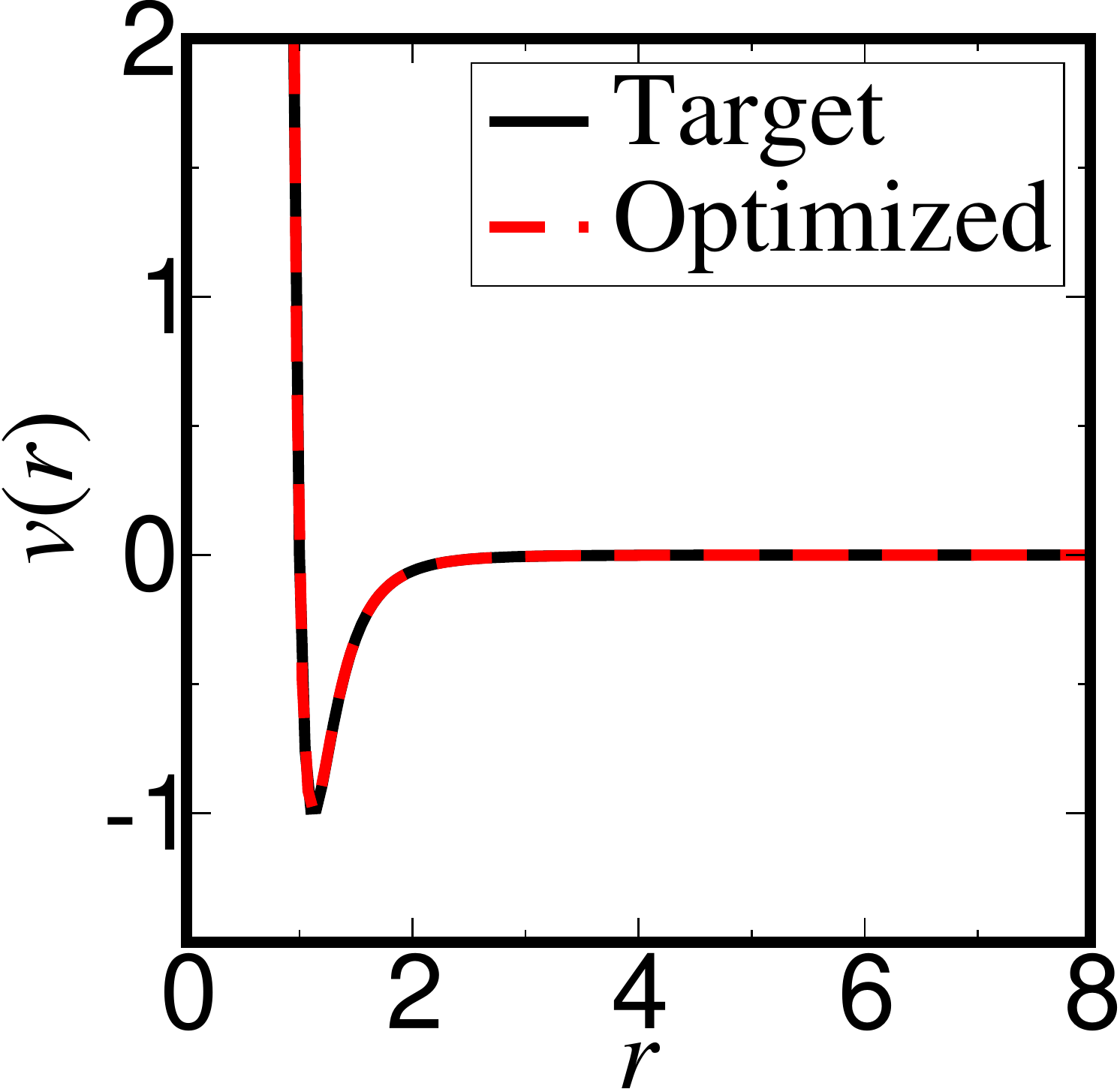}
}

\subfloat[]{
    \centering\includegraphics[width=6cm]{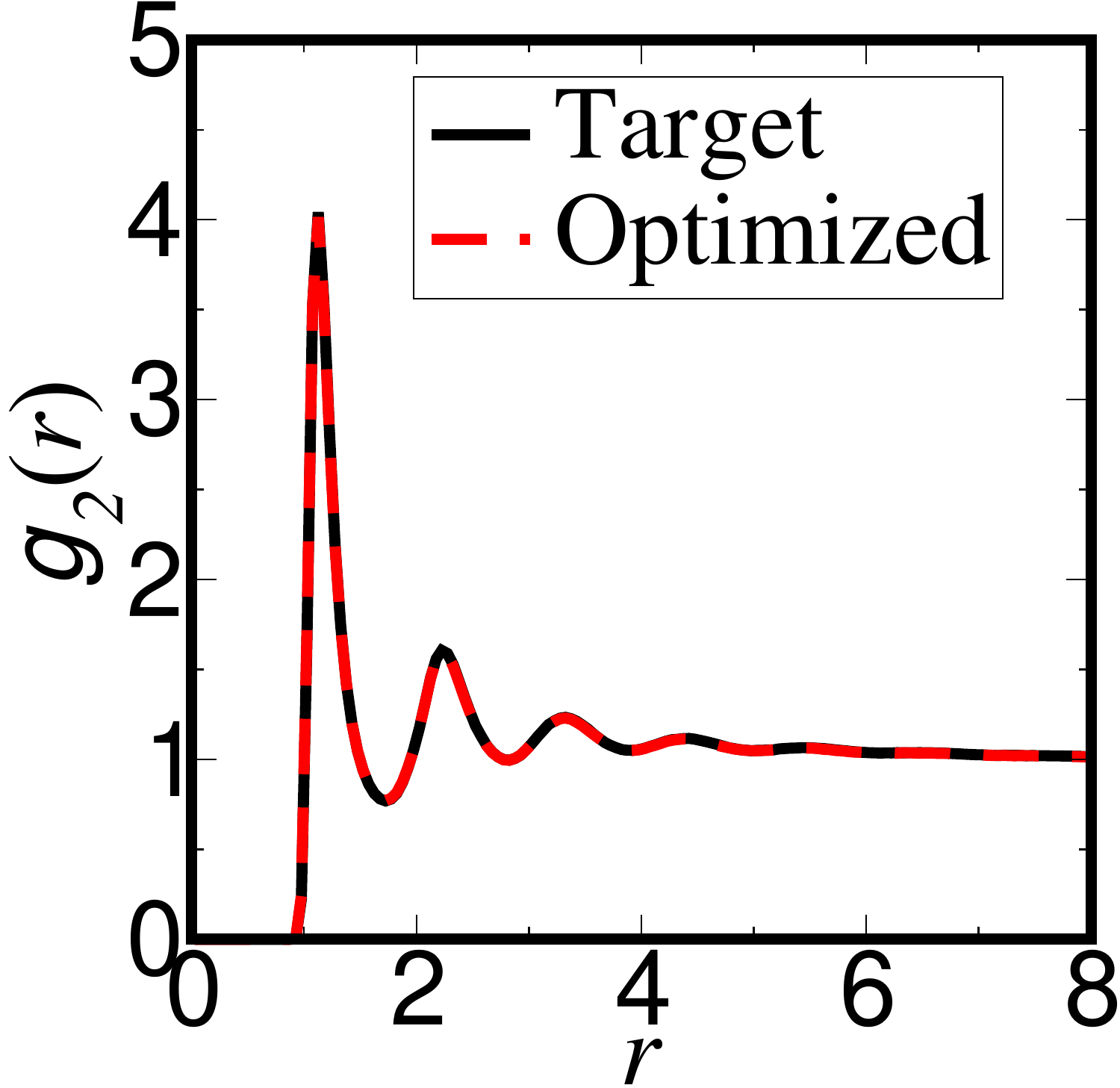}
}
\subfloat[]{
    \centering\includegraphics[width=5.9cm]{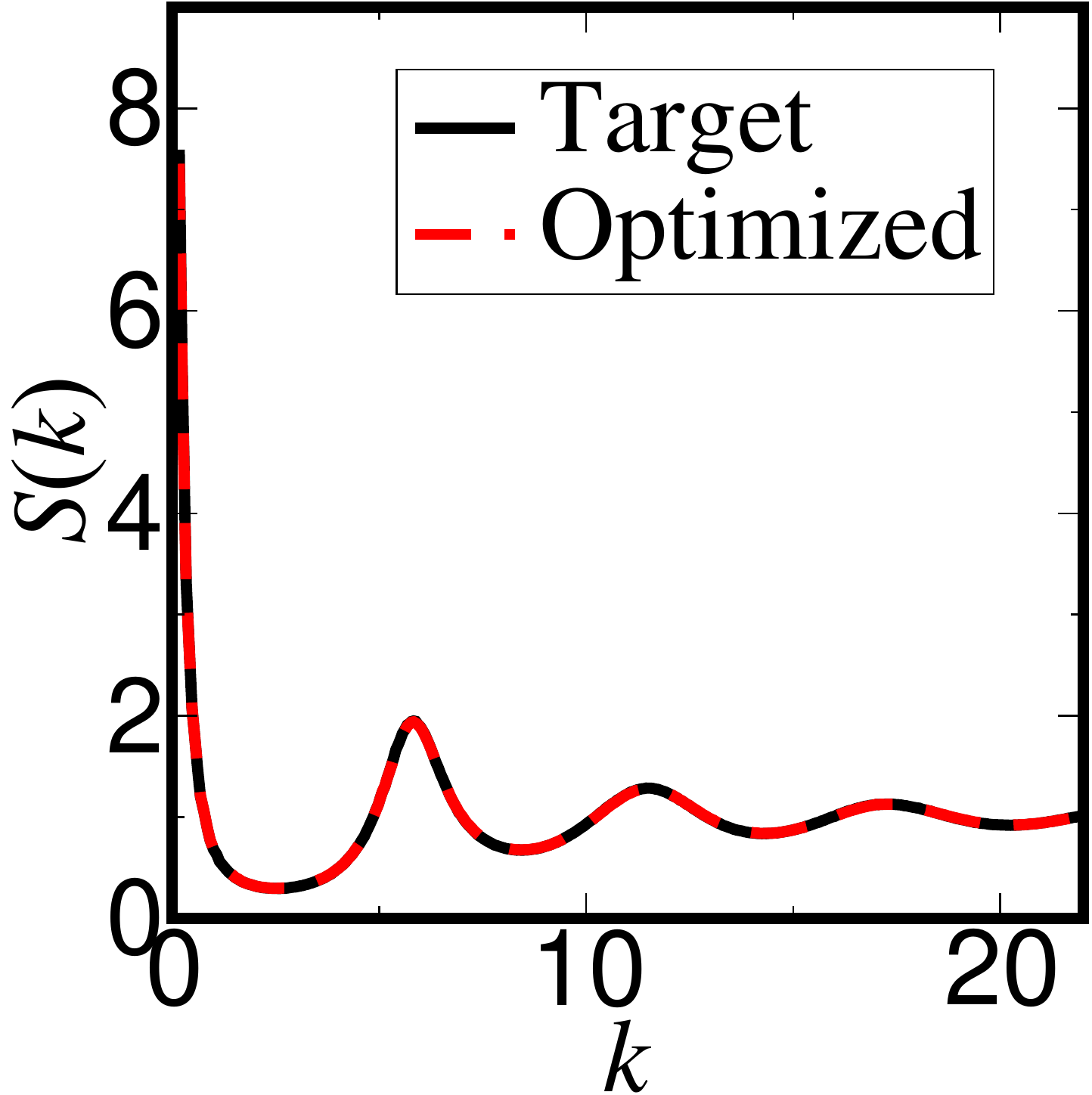}
}
\caption{(a) A snapshot of a 2D configuration 1,000-particle system that is equilibrated under the optimized potential [Eq. (\ref{ljGeneral})] for the target LJ fluid in the vicinity of its critical point. (b)  The target-generating potential $v_{\text{LJ}}(r)$ and the optimized pair potential $v_F(r;\mathbf{a})$. (c) Targeted and optimized pair correlation functions. (d) Targeted and optimized structure factors.}
  \label{fig:crit}
\end{figure*}


To apply our inverse methodology to systems in the vicinity of the liquid-gas critical point, which is nonhyperuniform, 
we generated a 2D systems equilibrated under the standard Lennard-Jones 6-12 potential 
\begin{equation}
    v_{\text{LJ}}(r)=4\varepsilon \left[ \left(\frac{\sigma}{r}\right)^{12} -\left(\frac{\sigma}{r}\right)^{6} \right]
    \label{lj}
\end{equation}
and then extracted the corresponding targeted pair statistics $g_{2,T}(r)$ and $S_T(k)$. The distance, density and temperature are made dimensionless as $r/\sigma$, $\rho\sigma^2$ and $k_BT/\varepsilon$.  Henceforth, we take 
$\varepsilon=1$, $\sigma=1$ and $k_B=1$.
We call $v_{\text{LJ}}(r)$ the \textit{target-generating} potential, which we know is unique according to Henderson's theorem \cite{He74}. Importantly, $v_{\text{LJ}}(r)$ was only used to generate the target pair statistics. When applying the inverse algorithm, we treat the target pair statistics simply as given, without regard to the target-generating potential, and hence do not assume the functional form of the potential. On the contrary, the initial guess of the basis functions of any potential is obtained via statistical-mechanical theory, and these are subject to re-selection during the optimization procedure; see Appendix for details. The density and temperature chosen for the target system are $\rho=\rho_c=0.37$ and  $T=1.1T_c=0.55$ \cite{Sm91,Li20}. A snapshot of the target system is shown in Fig. \ref{snapTarget}(a). This state point was chosen so that the critical scaling of the total correlation function (\ref{crit_h}) could be observed. We computed the  correlation length to be $\xi=4.83$ and fitted the Lorentzian form (\ref{lorentz}) for the structure factor $S_T(k)$ in the range $0<k<0.5$ to find $S_T(0)=12.2$, which is clearly nonhyperuniform. We further remark that the effect of critical slowing down \cite{Bi92} is significant for simulations at this state point, requiring a  relaxation time is two orders of magnitude larger than that for a typical dense Lennard-Jones liquid.

To apply our methodology described in Sec. \ref{meth}, we first accurately obtained the small-$k$ behavior for $\tilde{c}_T(k)$ and $S_T(k)$ via the Ornstein-Zernike equation (\ref{OZ}) as described in Sec. \ref{large-r}, and then performed inverse Fourier transform (\ref{inverse_fourier_radial}) to determine the large-$r$ behavior of $c_T(r)$, which we found to be an inverse power-law $-c_T(r)/\beta \sim v(r)\sim -Er^p$ [Eq. (\ref{power-law})]. Assuming that $p$ takes integer values, we fitted $\tilde{c}_T(k)$ with the form of Eq. (\ref{s_smallk}), but with $k^\zeta$ replaced by $\ln(k)k^\zeta$ when $\zeta$ is even, from which we robustly determined that $p=6$, implying $\zeta=4$.
To determine the form of the small-$r$ behavior of $v(r;\mathbf{a})$, we observed that $v_{\text{HNC}}(r)$ is strongly repulsive at small $r$, which is consistent with the inverse power-law form (\ref{power_law}) with $a_j^{(1)}>0$. Therefore,
the initial form of the pair potential was chosen to be
\begin{equation}
    v(r;\mathbf{a})=\varepsilon_1 r^{-p_1}-\varepsilon_2 r^{-6}.
    \label{ljGeneral}
\end{equation}
Note that although the form (\ref{ljGeneral}) is a generalized LJ potential, it was determined via statistical-mechanical theory from the target pair statistics alone, without knowledge of the functional form of the target-generating potential.
We verified that $v(r;\mathbf{a})$ satisfies the condition  $h_T^2(r)\ll|\beta v(r;\mathbf{a})|$ for large $r$, which is required to apply the asymptotic formula (\ref{asymp}). The reader is refereed to Appendix for additional implementation details.
Minimization of $\Psi$ using the form (\ref{ljGeneral}) yielded $\Psi<0.002$ within one single stage of optimization, i.e. no re-selection of the basis function was needed. The optimized parameters are given by $\varepsilon_1=3.98, p_1=11.93, \varepsilon_2=4.00$, which is in excellent agreement with the target-generating Lennard-Jones potential (\ref{lj}), as shown in Fig. \ref{fig:crit}(b). 
One can see that the configuration of the optimized equilibrium system [Fig. \ref{fig:crit}(a)] resembles that of the near-critical target  [Fig. \ref{snapTarget}(a)], and exhibits an expected large correlation length. As anticipated, there is excellent agreement between the target and optimized pair statistics in both direct and Fourier space [Fig. \ref{fig:crit}(c)--(d)]. 
The individual errors, defined by (\ref{g2-norm}) and (\ref{S-norm}), are given by $D_{g_2}=6.0\times 10^{-4}$ and $D_{S}=5.7\times 10^{-4}$.
The total $L_2$-norm error, defined by (\ref{L2}),  is  $\mathcal{E}=0.034$.

We also applied the IBI and the IHNCI procedures to the same target pair statistics. The trial potentials in the IBI procedure failed to converge and gave liquid-gas phase separated states.
The IHNCI procedure yielded a potential that is similar to our optimized potential but only in the small-$r$ range $0\leq r \leq 1$. However, it does not very accurately capture the large-$r$ behavior
of $v_{LJ}(r)$. The errors for the IHNCI procedure are $D_{g_2}=0.0074, D_{S}=0.011, \mathcal{E}=0.13$. Note that $\mathcal{E}$ obtained via IHNCI is an order
of magnitude larger than that obtained via our method due to the inaccuracy in the large-$r$ behavior of $v(r)$ as well as the accumulation of simulation errors in the binned potential. We also applied our methodology to LJ fluids at other state points, and find that for systems with long-ranged $g_2(r)$, including dense liquids near freezing, $\mathcal{E}$ obtained via our method is generally an order of magnitude lower than than that obtained via IHNCI. Therefore, our methodology appears to be superior to previous methods in solving these inverse problems for fluids in the critical region and dense liquids.

\subsection{Liquid under the Dzugutov potential}
\label{dzu}
\begin{figure*}[!ht]
\subfloat[]{
    \centering\includegraphics[width=5.5cm]{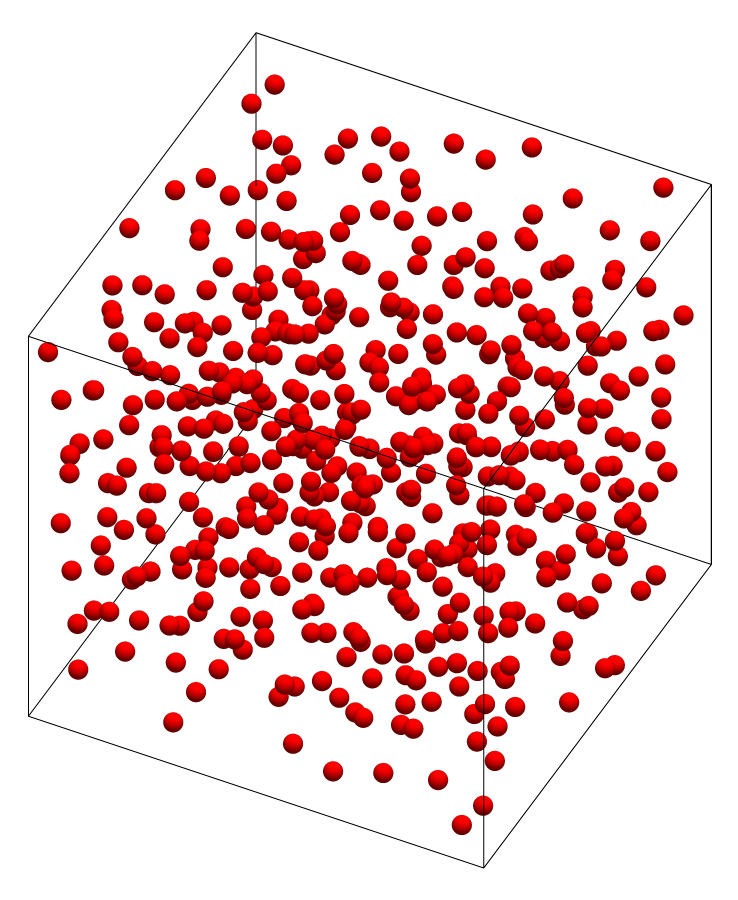}
}
\subfloat[]{
    \centering\includegraphics[width=6cm]{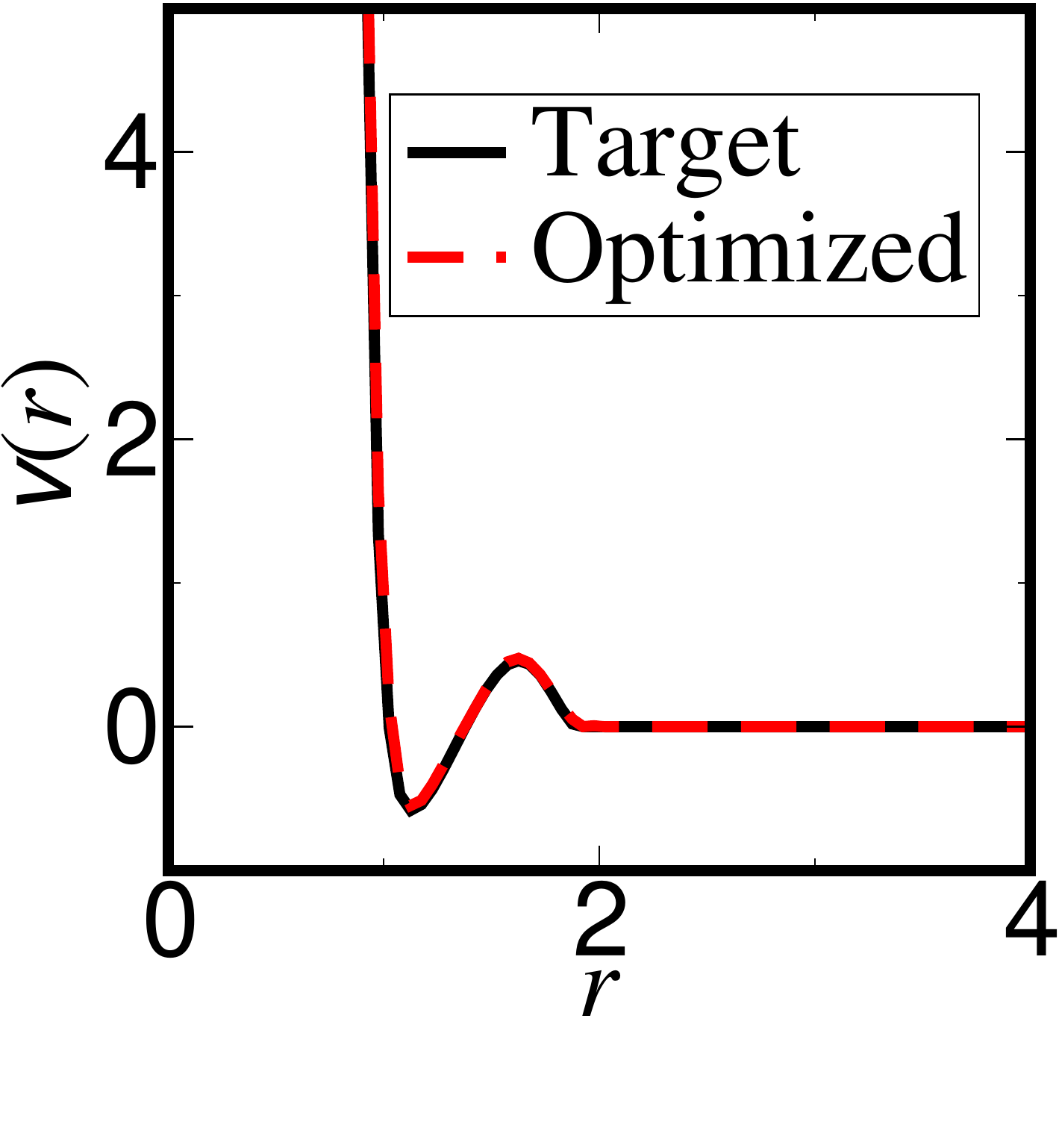}
}

\subfloat[]{
    \centering\includegraphics[width=6cm]{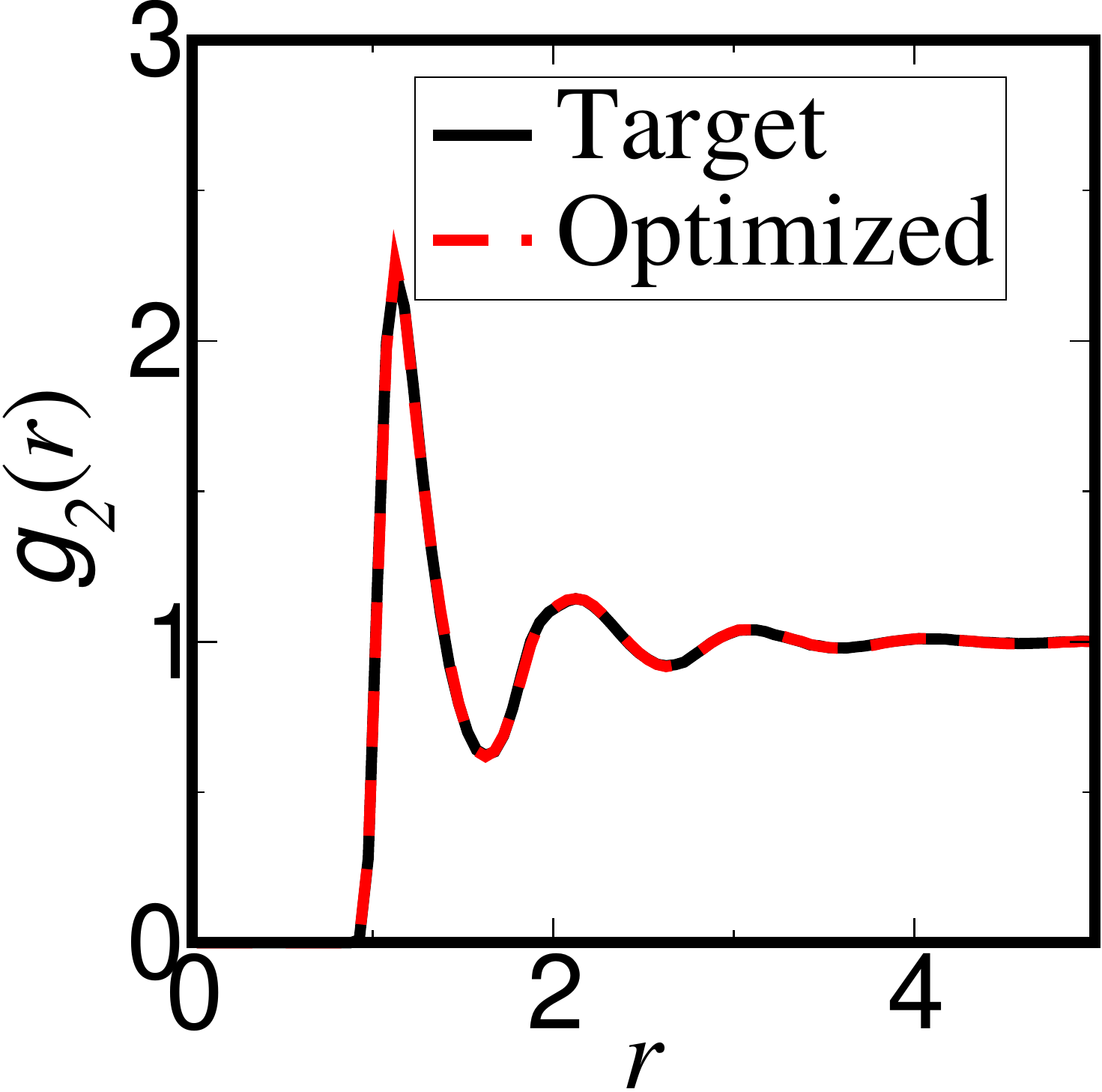}
}
\subfloat[]{
    \centering\includegraphics[width=5.9cm]{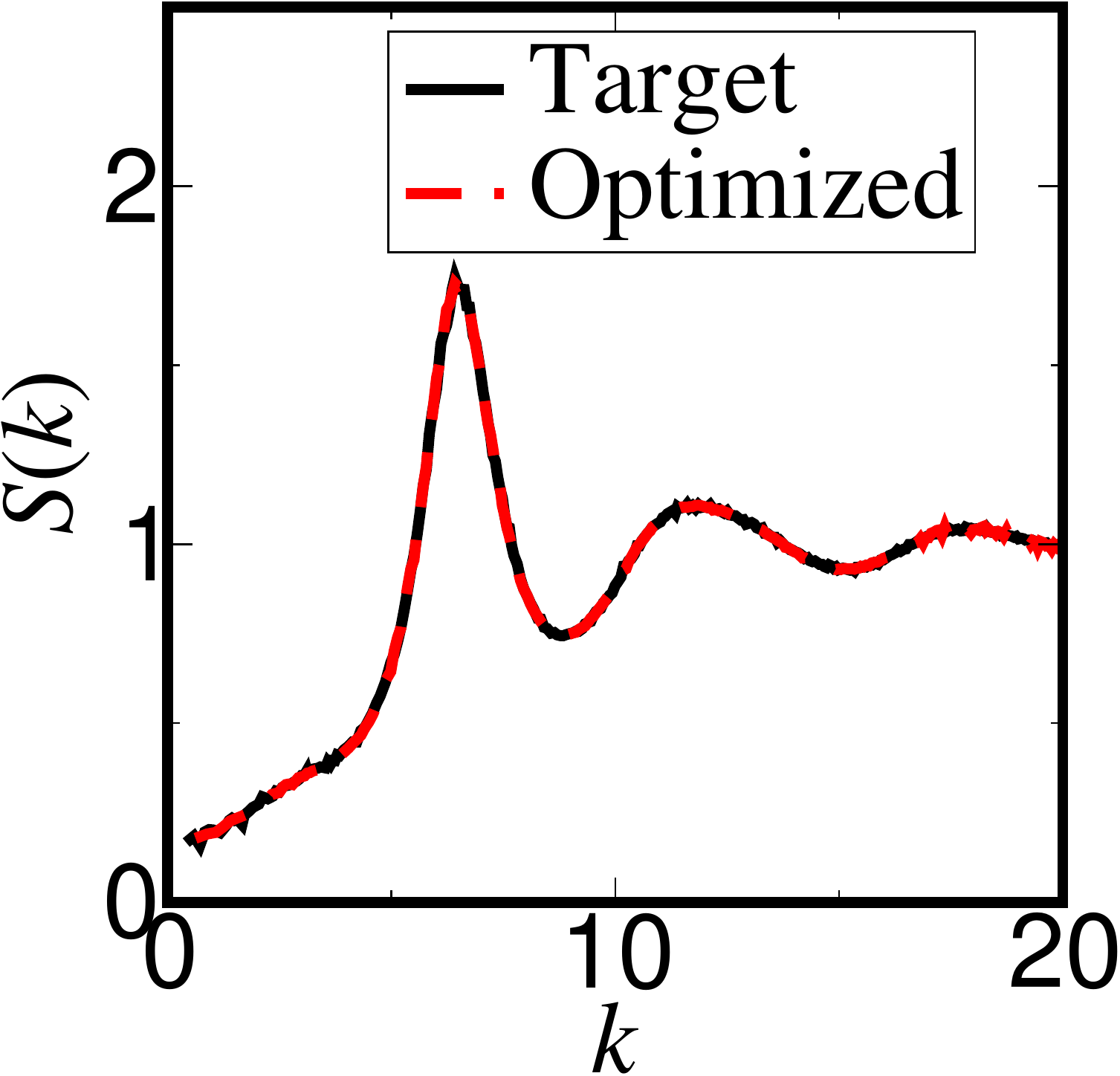}
}
\caption{(a) A snapshot of a 3D configuration 512-particle system that is equilibrated under the optimized potential [Eq. (\ref{dzu_v_opt})] for the Dzugutov liquid. (b)  The target-generating potential $v_{\text{D}}(r)$ (\ref{Dzu_def}) and the optimized pair potential (\ref{dzu_v_opt}). (c) Targeted and optimized pair correlation functions. (d) Targeted and optimized structure factors.}
  \label{fig:dzu}
\end{figure*}

To demonstrate that our inverse methodology is capable of treating systems equilibrated under nonstandard complex interactions, we study a 3D liquid under the Dzugutov potential, which is believed to be a good glass former \cite{Dz92, Dz03,Ma13a}. Following Ref. \cite{Dz92}, the form of the potential $v_{\text{D}}(r)$ is given by
\begin{widetext}
\begin{equation}
    \frac{v_{\text{D}}(r)}{\varepsilon}=\frac{A}{\varepsilon}\left[\left(\frac{r}{\sigma}\right)^{-m}-\frac{B}{\varepsilon}\right]\exp\left(\frac{c}{r - a}\right)H\left( \frac{a-r}{\sigma}\right) 
    + \frac{B}{\varepsilon}\exp(\frac{d}{r - b})H\left(\frac{b-r}{\sigma}\right),
    \label{Dzu_def}
\end{equation}
\end{widetext}
where $H(x)$ is the Heaviside step function, $A, B$ are energy parameters in units of $\varepsilon$, $m$ is a dimensionless exponent, $a, b, c, d$ are distance parameters in units of $\sigma$. The distance, density and temperature are made dimensionless as $r/\sigma$, $\rho\sigma^3$ and $k_BT/\varepsilon$. Henceforth, we take $\varepsilon=1$, $\sigma=1$ and $k_B=1$. The values of $A, B, m, a, b, c, d$ chosen for this study are given in the Appendix. 
Fig. \ref{fig:dzu}(b) plots the form of the potential, from which it is clear that $v_{\text{D}}(r)$ is piecewise smooth, with oscillations at small r and a cutoff at which the potential exhibits a discontinuous first derivative at the cutoff distance $r=b$. We generated a 3D systems equilibrated under $v_{D}(r)$ and then extracted the corresponding targeted pair statistics $g_{2,T}(r)$ and $S_T(k)$. The density and temperature chosen for the target system are $\rho=0.5$ and  $T=1$, which is a dense liquid state. Again, as in Sec. \ref{crit}, we do not assume the functional form (\ref{Dzu_def}) when applying the inverse algorithm, and the only input data to the algorithm are $\rho, T, g_{2,T}(r)$ and $S_T(k)$.

To apply our methodology described in Sec. \ref{meth}, we first accurately obtained the small-$k$ behavior for $\tilde{c}_T(k)$ from $S(k)$ via the Ornstein-Zernike equation (\ref{OZ}) and performed its inverse Fourier transform (\ref{inverse_fourier_radial}), from which we determined the large-$r$ behavior of $c_T(r)\sim -\beta v(r)$ has an  effectively superexponential decay. To obtain an initial guess of $v(r;\mathbf{a})$ for small- and intermediate-$r$, we observe that $v_{\text{HNC}}(r)$ has a strong repulsion for $r<1$ and is oscillatory in the range $1<r<2$. We fitted $v_{\text{HNC}}(r)$ with the forms \ref{gamma_form}--\ref{power_law_pot} and found the best fit of $v_{\text{HNC}}(r)$ is given by the exponential-damped oscillatory form \ref{exponential_form}. Thus, we started with an initial $v_(r;\mathbf{a})$ that is a sum of a power-law function and two exponential-damped basis functions. After one iteration of re-selecting basis functions, the final form of the optimized pair potential is given by
\begin{widetext}
\begin{equation}
    \frac{v_F(r;\mathbf{a})}{\varepsilon}=
    \begin{cases}
    \begin{split}
    &\sum_{i=1}^2\frac{\varepsilon_i}{r^{p_i}} 
    + \varepsilon_3\exp\left[-\left(\frac{r}{\sigma_3^{(1)}}\right)^2\right]\cos\left(\frac{r}{\sigma_3^{(2)}}+\theta_3\right) + \varepsilon_4\exp\left[-\left(\frac{r}{\sigma_4^{(1)}}\right)^4\right]\cos\left(\frac{r}{\sigma_4^{(2)}}+\theta_4\right)\\
    &+ \sum_{i=5}^7 \varepsilon_i\cos\left(\frac{r}{\sigma_i}+\theta_i\right)
    \end{split}
    \qquad r \leq 2\\
    \sum_{i=1}^2\frac{\varepsilon_i}{r^{p_i}} 
    + \varepsilon_3\exp\left[-\left(\frac{r}{\sigma_3^{(1)}}\right)^2\right]\cos\left(\frac{r}{\sigma_3^{(2)}}+\theta_3\right)
    + \varepsilon_4\exp\left[-\left(\frac{r}{\sigma_4^{(1)}}\right)^4\right]\cos\left(\frac{r}{\sigma_4^{(2)}}+\theta_4\right) \qquad r > 2,
    \end{cases}
\label{dzu_v_opt}
\end{equation}
\end{widetext}
where the undamped oscillatory functions $f_5,f_6,f_7$ resulted from the re-selection of the basis functions, which fine-tunes $v(r;\mathbf{a})$ in the range $0\leq r \leq 2$; see Appendix for details.
Note that while the functional form (\ref{dzu_v_opt}) is different from that of the target-generating potential (\ref{Dzu_def}), this is expected, because the latter is not known \textit{a priori} in application of our algorithm. However, Fig. \ref{fig:dzu}(b) show that the plots for $v_F(v;\mathbf{a})$ and $v_{\text{D}}(r)$ closely agree with one another, indicating that our methodology has precisely recovered the target-generating potential. We verified that $v_F(r;\mathbf{a})$ satisfies the condition  $h_T^2(r)\ll|\beta v(r;\mathbf{a})|$ for large $r$, which is required to apply the asymptotic formula (\ref{asymp}). The reader is refereed to Appendix for additional implementation details. Figure \ref{snapTarget}(a) and \ref{fig:dzu}(a) show that configurations of the target and optimized systems closely resemble each other. There is also excellent agreement between the target and optimized pair statistics in both direct and Fourier space [Fig. \ref{fig:dzu}(c)--(d)]. 
The individual errors, defined by (\ref{g2-norm}) and (\ref{S-norm}), are given by $D_{g_2}=0.0010$ and $D_{S}=0.0010$.
The total $L_2$-norm error, defined by (\ref{L2}),  is  $\mathcal{E}=0.045$. On the other hand, the errors for the IHNCI procedure is $D_{g_2}=0.0040, D_{S}=0.0049, \mathcal{E}=0.094$. Note that $\mathcal{E}$ obtained via IHNCI about twice of that obtained via our method due to the the accumulation of simulation errors in the binned potential.

\subsection{Equilibrium system corresponding to nonequilibrium random sequential addition pair statistics}
\label{rsa}

Our methodology provides a powerful means to test  the Zhang-Torquato conjecture \cite{Zh20}
by determining whether classical equilibrium systems  under an effective pair interaction
can be attained that accurately match the pair statistics of nonequilibrium systems. Probing systems with identical pair statistics but different higher-body statistics will also shed light on the degeneracy problem \cite{Ji10a,St19}.  


Here, we choose our nonequilibrium  target system to be the random sequential addition (RSA) packing process \cite{Wi66,Fe80,Co88,To06a,Zh13b},
 which has important applications in protein and polymer deposition models. Starting with an empty but large volume $\Omega$ in $\mathbb{R}^d$, the RSA process is produced by randomly, irreversibly, and sequentially placing nonoverlapping (hard) spheres of diameter $D$ into this volume
under periodic boundary conditions subject to a nonoverlap constraint. If a new sphere does not overlap with any existing spheres at some time $t$, it is added to the configuration; otherwise, the attempt is discarded. One can stop the addition process at any $t$, obtaining RSA configurations with a range of packing fractions $\phi(t)$ up to the infinite-time  maximal saturation value $\phi_s = \phi(\infty)$. For identical spheres in 2D, $\phi_s\approx 0.547$ \cite{Fe80,To06a,Zh13b}.

\begin{figure*}[htp]
  \centering
  \subfloat[]{\label{snap}\includegraphics[width=54mm]{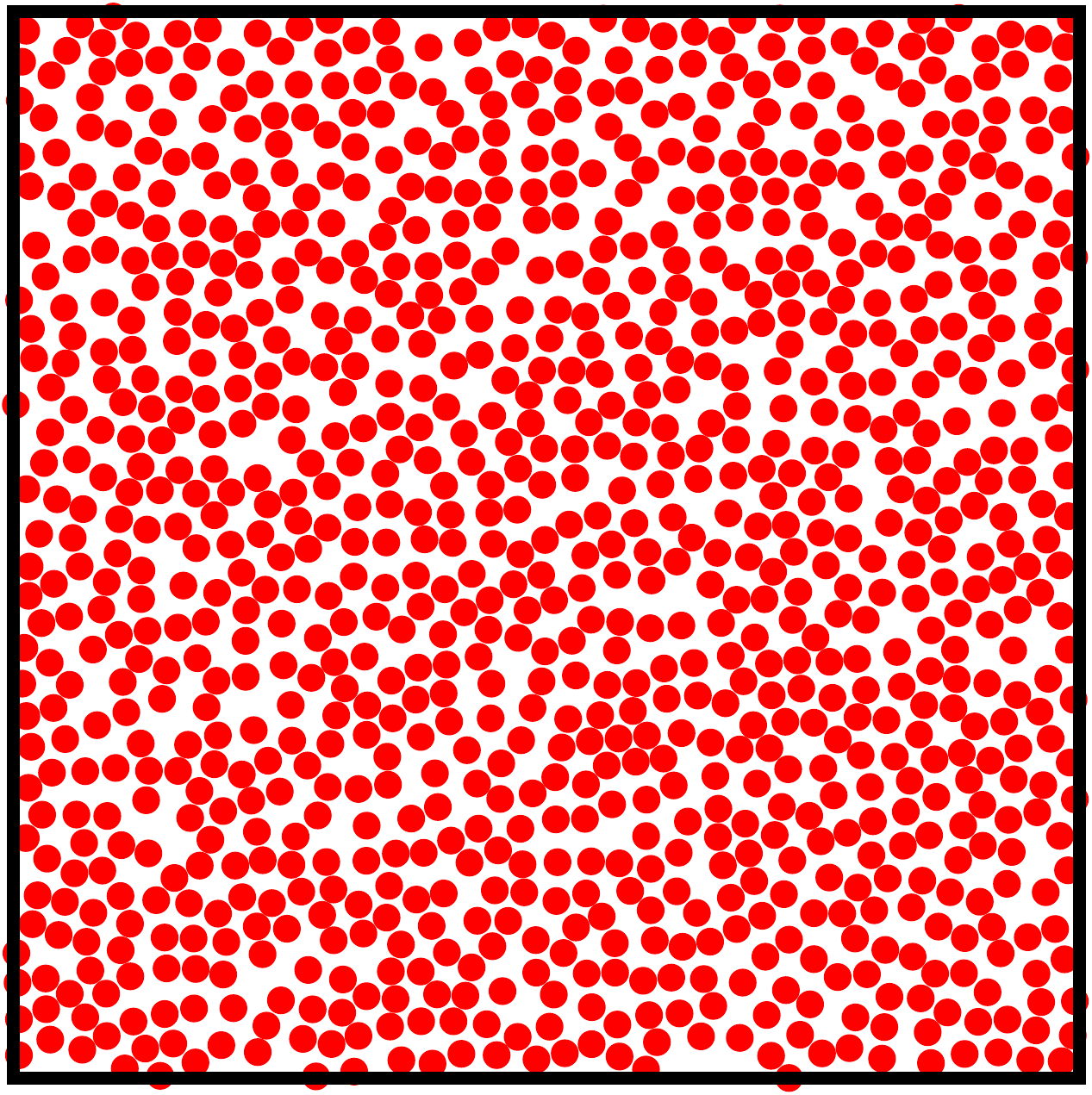}}
  \hspace{0.5em}
  \subfloat[]{\label{rsa2d_v}\includegraphics[width=62mm]{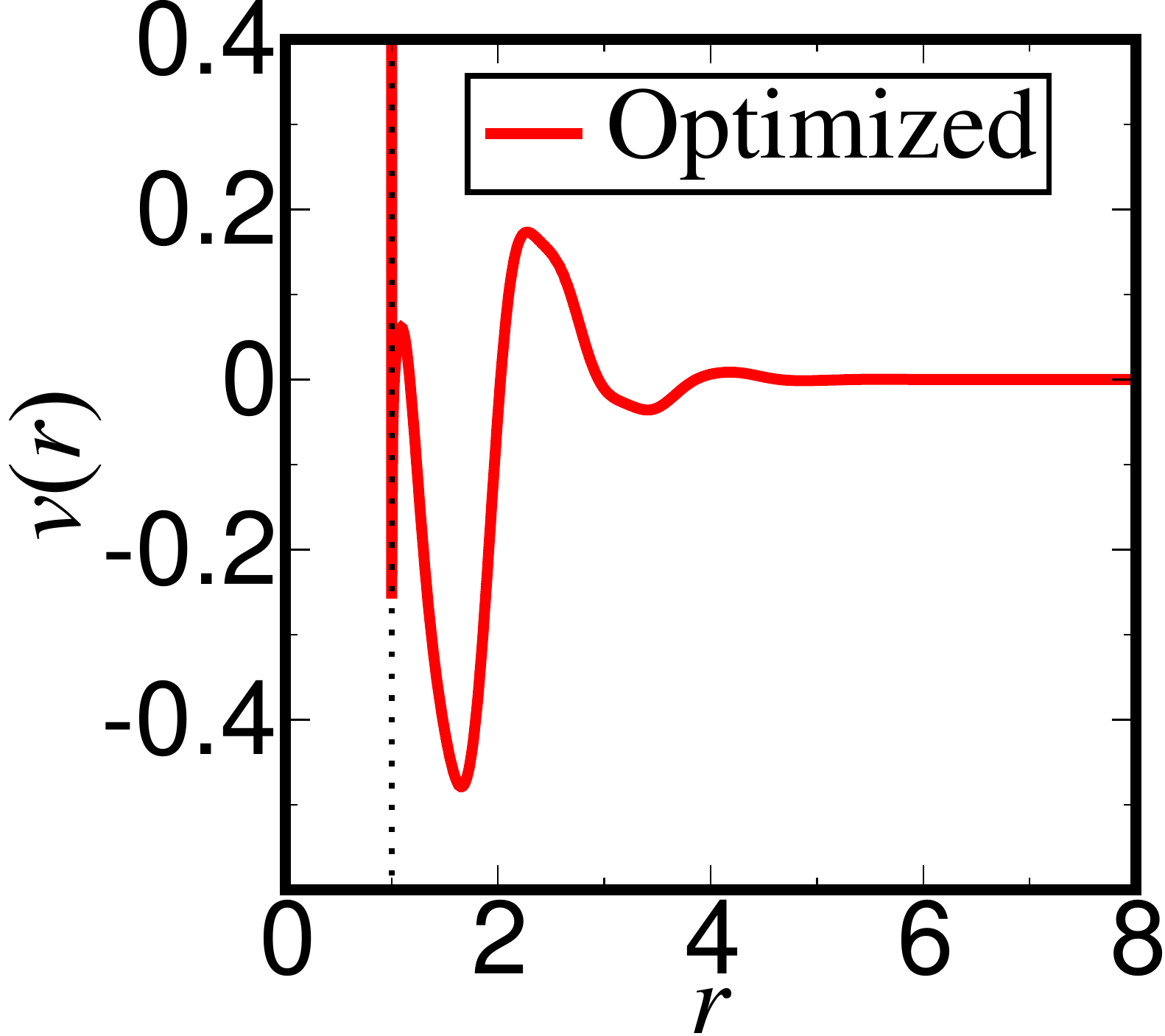}}
  \\
  \subfloat[]{\label{rsa2d_g2}\includegraphics[width=58mm]{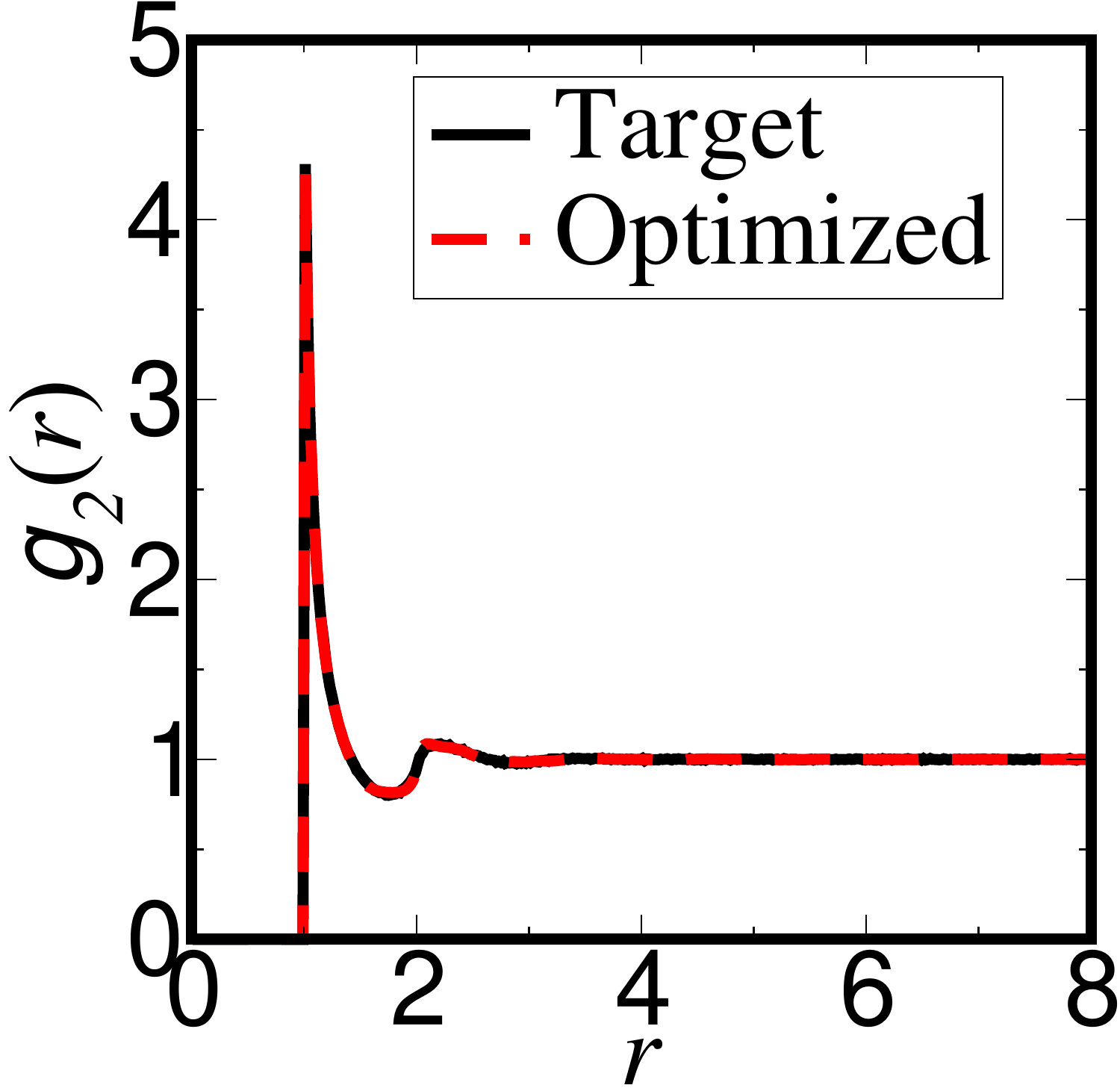}}
  \hspace{0.5em}
  \subfloat[]{\label{rsa2d_s}\includegraphics[width=60mm]{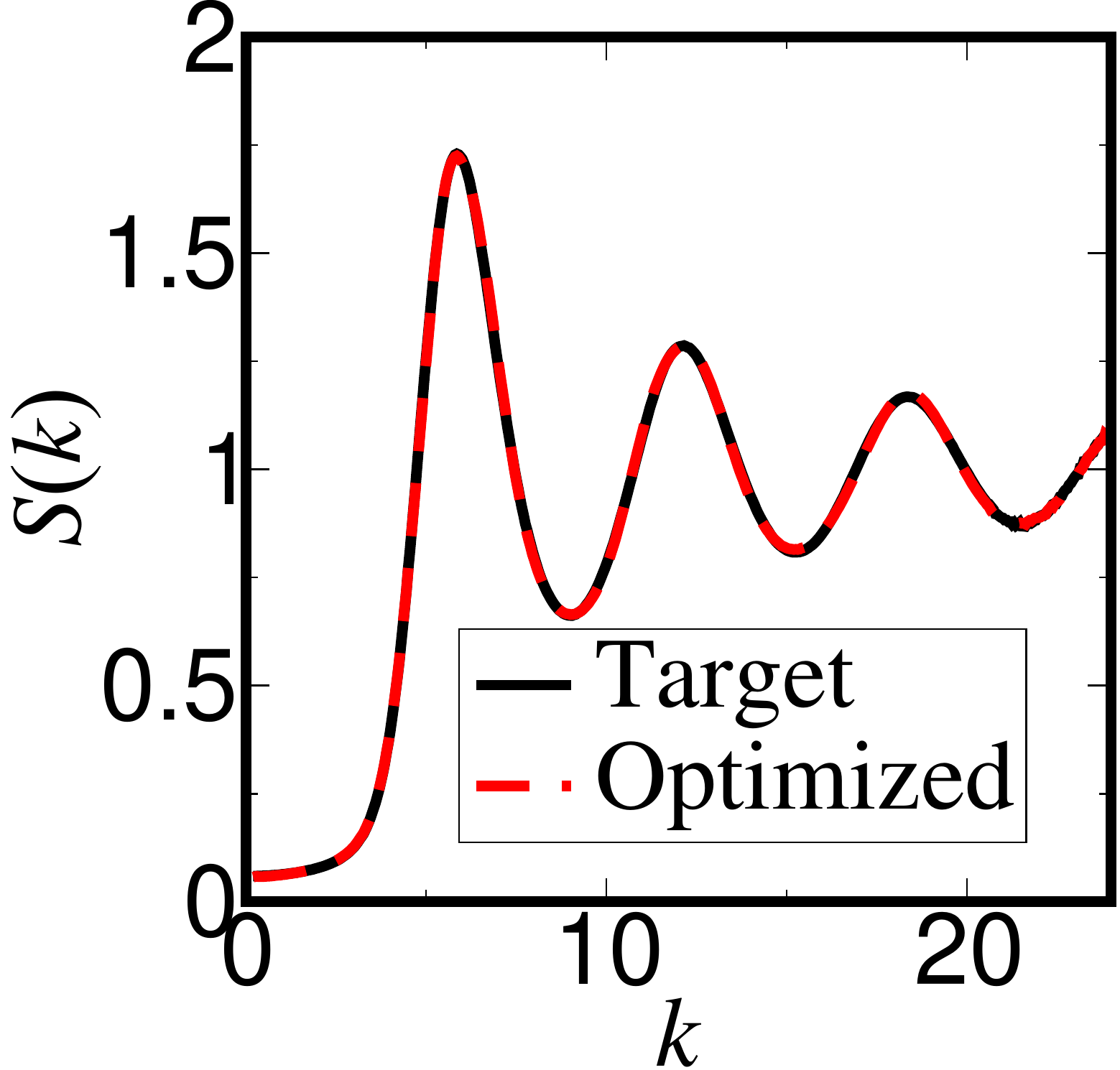}}
  \caption{(a) A snapshot a 2D configuration 1,000-particle   system that is equilibrated under the optimized potential [Eq. (\ref{rsapot})] for the 
target nonequilibrium RSA packing near its saturation state. (b) Optimized pair potential. (c) Targeted and optimized pair correlation functions. (d) Targeted and optimized structure factors.}
  \label{fig:rsa}
\end{figure*}

It is known that $g_2(r)$'s for RSA processes at the saturation limit possess a logarithmic divergence as $r\rightarrow D^+$, independent of dimension \cite{Po80}. This suggests that the pair potential corresponding to a saturated RSA packing, if it exists, also possesses  a corresponding
singularity. Therefore, we choose as our target a 2D RSA packing very near, but not exactly at, saturation, with $\phi = 0.534 = 0.976\phi_s$,
which avoids treating the aforementioned singularity, but is still a challenging target.  
We set $D$ to be unity to generate the target RSA system. A snapshot of the target system is given in Fig. \ref{fig:rsa}(a).

The effective pair potential $v(r;\mathbf{a})$ must have a hard core [Eq. (\ref{hard_core})] for $r\leq 1$ to respect
the nonoverlap constraint. To determine the large-$r$ behavior of $v(r;\mathbf{a})$, we used the asymptotic analysis of $\tilde{c}_{T}(k)$ described in Sec. \ref{large-r} and found that that $\beta c_{T}(r)$ has superexponential decay rate that is best described by the a Gamma-function form (\ref{gamma_form}). For 
intermediate values of $r$, we observed that the HNC approximation contains a minimum at the contact radius $r=1$ and exhibits apparent oscillations about the horizontal axis. Therefore, starting from an initial potential that is a sum of a hard core, a non-oscillatory Gamma-form function (for the minimum at contact), and two Gamma-damped oscillatory functions, we went through 4 iterations of optimization and re-choosing basis functions and arrived at the following final form of the parameterized pair potential:
\begin{widetext}
\begin{equation}
  v_F(r;\mathbf{a})=
\begin{cases}
    \infty \qquad &r\leq 1\\
    \varepsilon\left[-\sum_{j=1}^{2}\frac{\varepsilon_j}{\Gamma\left(r/\sigma_j\right)} + \frac{\sum_{j=3}^{7}\varepsilon_j\cos\left(r/\sigma_j^{(1)} + \theta_j\right)}{\Gamma
    \left(r/\sigma_3^{(2)}\right)}\right] \qquad &r>1,
\end{cases}
\label{rsapot}
\end{equation}
\end{widetext}
which satisfies the condition $h_T^2(r)\ll|\beta v_F(r;\mathbf{a})|$ for large $r$, which is required to apply the asymptotic formula (\ref{asymp}).
In our simulations, we let both the energy scale $\varepsilon$ and the dimensionless temperature $k_BT/\varepsilon$ to be unity.
The optimized parameters are provided in Table \ref{rsa_params} of the Appendix, which includes other implementations details. First, observe that the configuration of the optimized system Fig. \ref{fig:rsa}(a) is visually very similar to the targeted nonequilibrium configuration in Fig. \ref{snapTarget}(c). Figure \ref{fig:rsa}(b) shows the optimized potential, which possesses a sharp well at the hard-sphere diameter $r=1$, and another broader well at $r=1.66$. As shown in Fig. \ref{fig:rsa}(c)--(d), the optimized pair potential gives pair statistics that agree well with the target nonequilibrium pair statistics. 
The individual errors are $D_{g_2}=0.0023$ and $D_{S}=0.0015$ and the total $L_2$-norm error is $ \mathcal{E}=0.062$. This provides a vivid example that demonstrably adds to the growing evidence of the validity of the Zhang-Torquato conjecture \cite{Zh20}. 

By contrast, the individual errors resulting from the IHNCI procedure are $D_{g_2}=0.0033$ and $ D_{S}=0.0060$,
while the total $L_2$-norm errors is $\mathcal{E}=0.096$, which are slightly larger than those from our method due to the accumulation of random errors. The fact that the IHNCI is relatively accurate for this target is due to the superexponential decay of $g_{2,T}(r)$, which translates to a fast decaying pair potential $v(r)$ via the HNC approximation.

\subsection{Equilibrium system corresponding to nonequilibrium 3D cloaked URL system}
\label{url}
In this subsection, we apply our methodology to the challenging task of determining an equilibrium system corresponding to the pair statistics of a nonequilibrium hyperuniform {\it cloaked} \textit{uniform randomized lattice (URL) model \cite{Kl20}}. In URL models, each point in a $d$-dimensional simple cubic lattice $\mathcal{L} = \mathbb{Z}^d$ is displaced by a random vector that is uniformly distributed on a rescaled unit cell $bC = [-b/2, b/2)^d$ , where $b > 0$ is a scalar factor and $C$ is a unit cell of the lattice. By definition, the lattice constant is set to be unity. It has been shown that the structure factor for the URL point process contains Bragg peaks that coincide with the unperturbed lattice as well as a diffuse part such that $\lim_{k\rightarrow 0}S(k)\sim k^2$ \cite{Ga04,Ki18a,Kl20}. Remarkably, Klatt et al. showed that the Bragg peaks in the structure factors vanish completely, or become ``cloaked'', when $b$ takes integer values \cite{Kl20}. 

We applied our methodology to determine a positive-temperature equilibrium system with one- and two-body interactions that realizes the pair statistics of a 3D cloaked URL model with $\rho=1, b=1$. Such a classical system must be thermodynamically incompressible; see Sec. \ref{hyperuniformEquilStates}. The fact that the target system is hyperuniform with $\alpha=2$ implies that the effective pair potential has the asymptotic form $v(r)\sim 1/r$; see Eq. (\ref{v_hu}). After fitting the HNC approximation with Eqs. (\ref{gamma_form})--(\ref{power_law}) for the small-$r$ range $0\leq r \leq 2$, we determined the initial form of the pair potential to be 
\begin{equation}
   v(r;\mathbf{a})=\varepsilon\left[v_l(r;\mathbf{a}_l) + v_s(r;\mathbf{a}_s)\right],
\label{url_v}
\end{equation}
where 
\begin{equation}
    v_l(r;\mathbf{a}_l) = \frac{\varepsilon_1}{r}
\end{equation}
and 
\begin{equation}
    v_s(r;\mathbf{a}_s) = -\frac{\varepsilon_2\exp\left(-r/\sigma_2\right)}{r^{p_2}}+\varepsilon_3 \exp\left[- \left(\frac{r-\sigma_3^{(1)}}{\sigma_3^{(2)}}\right)^2\right]
\end{equation}
are the long-ranged and short-ranged parts of $v(r;\mathbf{a})$, respectively; see the Appendix for details. We verified that Eq. (\ref{url_v}) satisfies the condition $h_T^2(r)\ll|\beta v(r;\mathbf{a})|$ for large $r$ to apply (\ref{asymp}). Henceforth, we set both the energy scale $\varepsilon$ and the dimensionless temperature $k_BT/\varepsilon$ to be unity.
We reiterate that $v_l(r;\mathbf{a}_l)$ can be regarded as a Coulombic interaction between ``like-charged'' particles. The one-body potential was treated using the procedure described in Sec. \ref{meth_hu}. Under periodic boundary conditions with a cubic simulation box of edge length $L$, the effective interaction between particle $i$ and all images of particle $j$ (\ref{hu_vpbc}) is given by
\begin{widetext}
\begin{equation}
    v_{e,\text{PBC}}(\mathbf{r}_{ij};\mathbf{a})=\sum_{\mathbf{n}}\frac{\varepsilon_1}{|\mathbf{r}_{ij}+\mathbf{n}L|}
    -\frac{\varepsilon_1}{L^3}\int_{\mathbb{R}^3}\frac{1}{x}d\mathbf{x}
    +\sum_{\mathbf{n}} v_s(|\mathbf{r}_{ij}+\mathbf{n}L|;\mathbf{a}_s)
    - \frac{1}{L^3}\int_{\mathbb{R}^3}v_s(x;\mathbf{a}_s)d\mathbf{x}
    \label{vpbc}
\end{equation}
\end{widetext}
where $\mathbf{n}$ are sites of the cubic lattice $\mathbb{Z}^3$. Since $v_s(r;\mathbf{a}_s)$ is short-ranged, the third term in (\ref{vpbc}) is dominated by the minimum image of $j$ and the fourth term of (\ref{vpbc}) vanishes in the limit of large $L$. Therefore,
\begin{widetext}
\begin{equation}
     v_{e,\text{PBC}}(\mathbf{r}_{ij};\mathbf{a})=
     \sum_{\mathbf{n}}\frac{\varepsilon_1}{|\mathbf{r}_{ij}+\mathbf{n}L|}
    -\frac{\varepsilon_1}{L^3}\int_{\mathbb{R}^3}\frac{1}{x}d\mathbf{x} + v_s(r_{ij}), \qquad L\rightarrow\infty.
    \label{vpbc_largeL}
\end{equation}
\end{widetext}

The first two terms of (\ref{vpbc_largeL}), which are integrals over long-ranged potentials, can be converted to an absolutely convergent integral and efficiently evaluated using the Ewald summation technique \cite{Ew21,Ha73}:
\begin{widetext}
\begin{equation}
\begin{split}
   &\sum_{\mathbf{n}}\frac{\varepsilon_1}{|\mathbf{r}_{ij}+\mathbf{n}L|}-\frac{\varepsilon_1}{L^3}\int_{\mathbb{R}^3}\frac{1}{x}d\mathbf{x}=\sum_{\mathbf{n}}\frac{\varepsilon_1}{|\mathbf{r}_{ij}+\mathbf{n}L|}-\frac{\varepsilon_1}{L^3}\int_{\mathbb{R}^3}\frac{1}{|\mathbf{r}_{ij}+\mathbf{x}|}d\mathbf{x}\\
  &=\frac{\varepsilon_1}{L}\left(\sum_{\mathbf{n}}\frac{\text{erfc}(\pi^{1/2}|\mathbf{r}_{ij}/L+\mathbf{n}|)}{|\mathbf{r}_{ij}/L+\mathbf{n}|}-1+\sum_{\mathbf{n}\ne \mathbf{0}}\frac{1}{\pi n^2}\exp(-\pi n^2)\exp[2i\pi\mathbf{n}(\mathbf{r}_{ij}/L)]\right), 
   \end{split}
    \label{ewald}
\end{equation}
where $n=|\mathbf{n}|$. Thus,
\begin{equation}
\begin{split}
    v_{e,\text{PBC}}(\mathbf{r}_{ij};\mathbf{a})&=\frac{\varepsilon_1}{L}\left(\sum_{\mathbf{n}}\frac{\text{erfc}(\pi^{1/2}|\mathbf{r}_{ij}/L+\mathbf{n}|)}{|\mathbf{r}_{ij}/L+\mathbf{n}|}-1+\sum_{\mathbf{n}\ne \mathbf{0}}\frac{1}{\pi\lambda^2}\exp(-\pi\lambda^2)\exp(2i\pi\mathbf{n}(\mathbf{r}_{ij}/L))\right)\\ &-\frac{\varepsilon_2\exp\left(-r/\sigma_2\right)}{r^{p_2}}+\varepsilon_3 \exp\left[- \left(\frac{r-\sigma_3^{(1)}}{\sigma_3^{(2)}}\right)^2\right], \qquad L\rightarrow \infty
    \end{split}
    \label{vpbc_full}
\end{equation}
\end{widetext}
which is the effective potential that we used in Monte Carlo simulations under periodic boundary conditions.

The form of the parameterized potential Eq. (\ref{url_v}) achieved the desired convergence criterion $\Psi<0.002$ in one single optimization stage, i.e., no re-selection of basis function was needed. The optimized parameters are listed in Table \ref{URL_params} of the Appendix. Figures \ref{snapTarget}(d) and \ref{3DURL_basic}(a) show the snapshots of the target and the optimized systems, respectively, which are visually indistinguishable. Figure \ref{3DURL_basic}(b) plots the short-ranged part of the effective potential, i.e. $v_s(r;\mathbf{a}_s)=v(r;\mathbf{a})-0.940/r$ against $r$. Figure \ref{3DURL_basic}(c) and (d) show $g_2(r)$ and $S(k)$, respectively, for the target and optimized systems. The
 pair statistics of the target cloaked URL is in excellent agreement with those of the optimized equilibrium system in both direct and Fourier space. The individual
errors are $D_{g_2}=4.8\times 10^{-4}$ and  $D_{S}=7.5\times 10^{-4}$ and the $L_2$-norm error is $\mathcal{E}=0.035$,
all of which are remarkably relatively small.

Importantly, the IBI and IHNCI procedures cannot treat hyperuniform targets. First, they do not include one-body potentials. Second, they impose a cut-off on the extent of the pair potential, which is not appropriate for the required long-ranged interactions.

\begin{figure*}[htp]
  \centering
  \subfloat[]{\label{3DURL_snapInferred}\includegraphics[width=54mm]{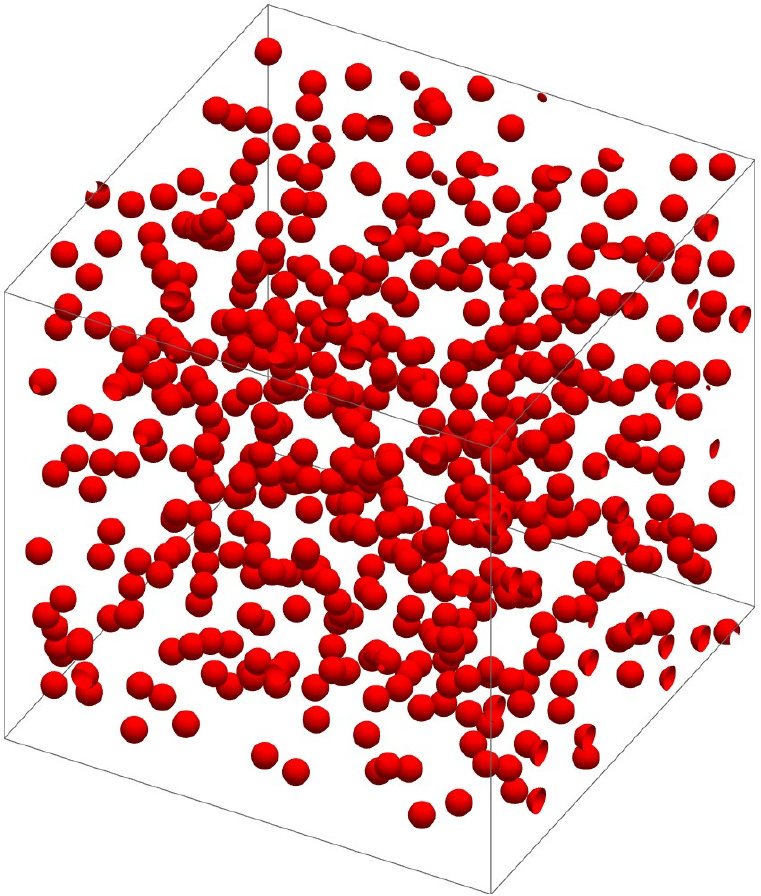}}
  \hspace{0.5em}
  \subfloat[]{\label{3DURL_v}\includegraphics[width=62mm]{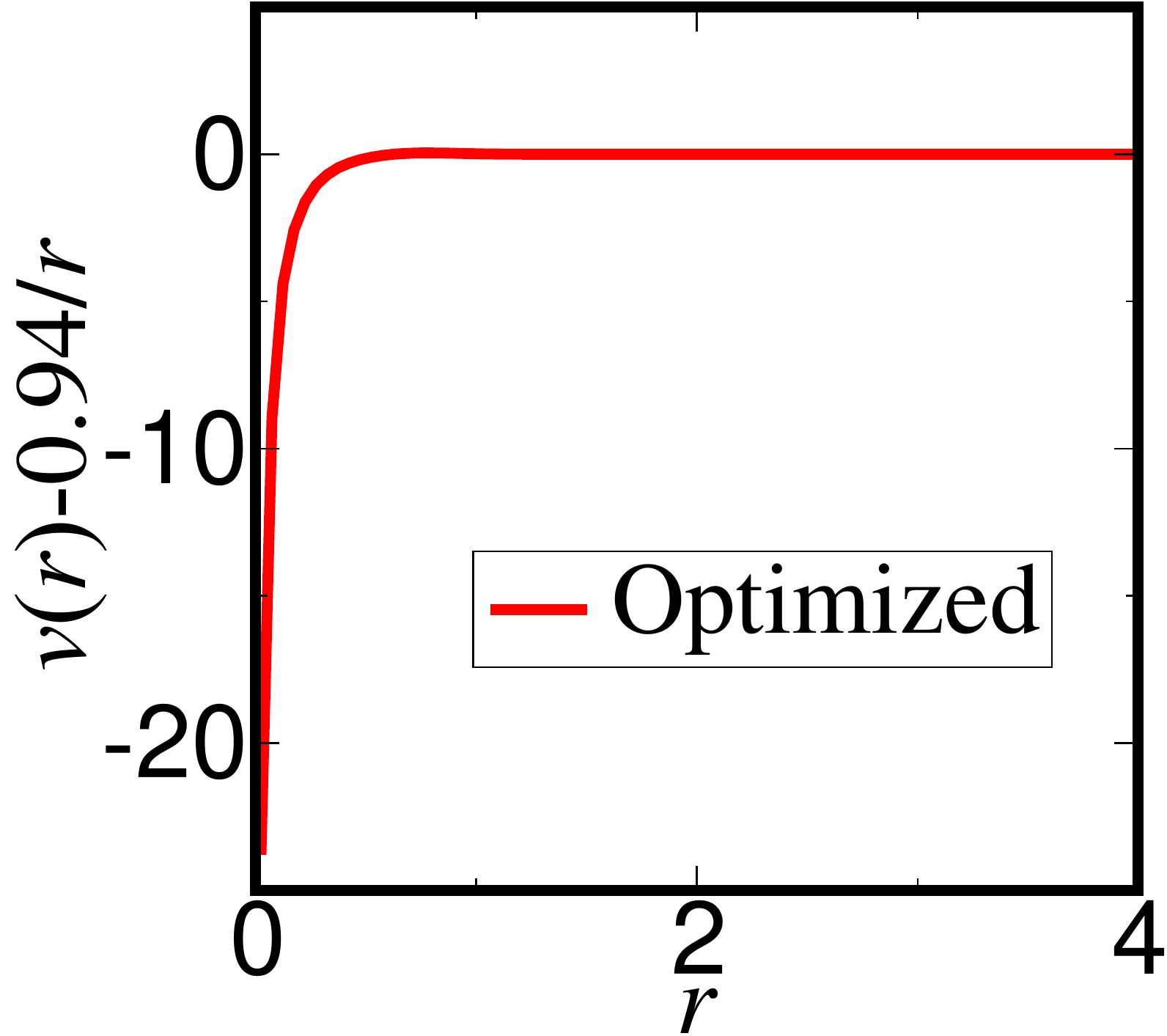}}
    \\
  \subfloat[]{\label{3DURL_g2}\includegraphics[width=58mm]{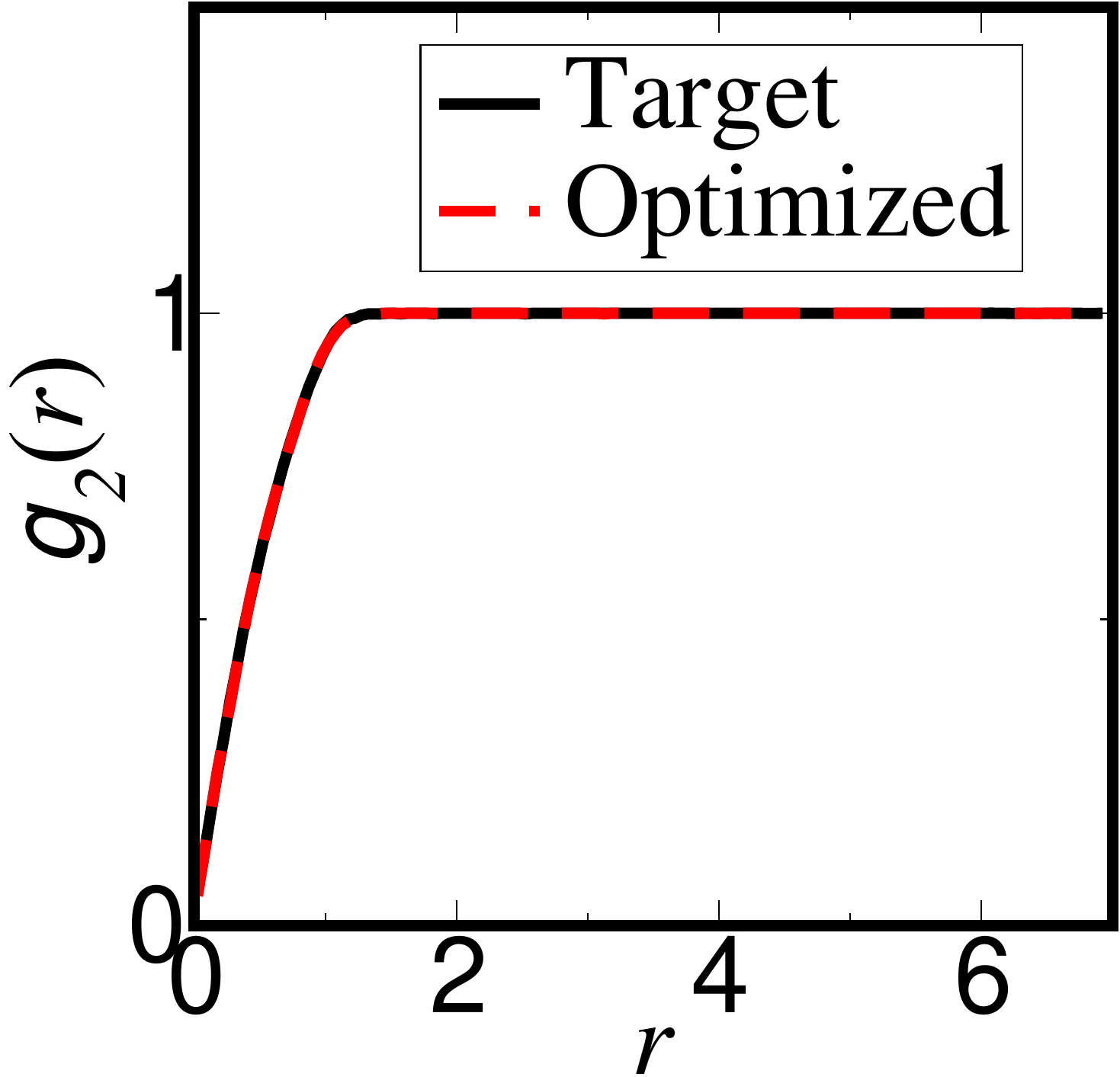}}
  \hspace{0.5em}
  \subfloat[]{\label{3DURL_S}\includegraphics[width=57mm]{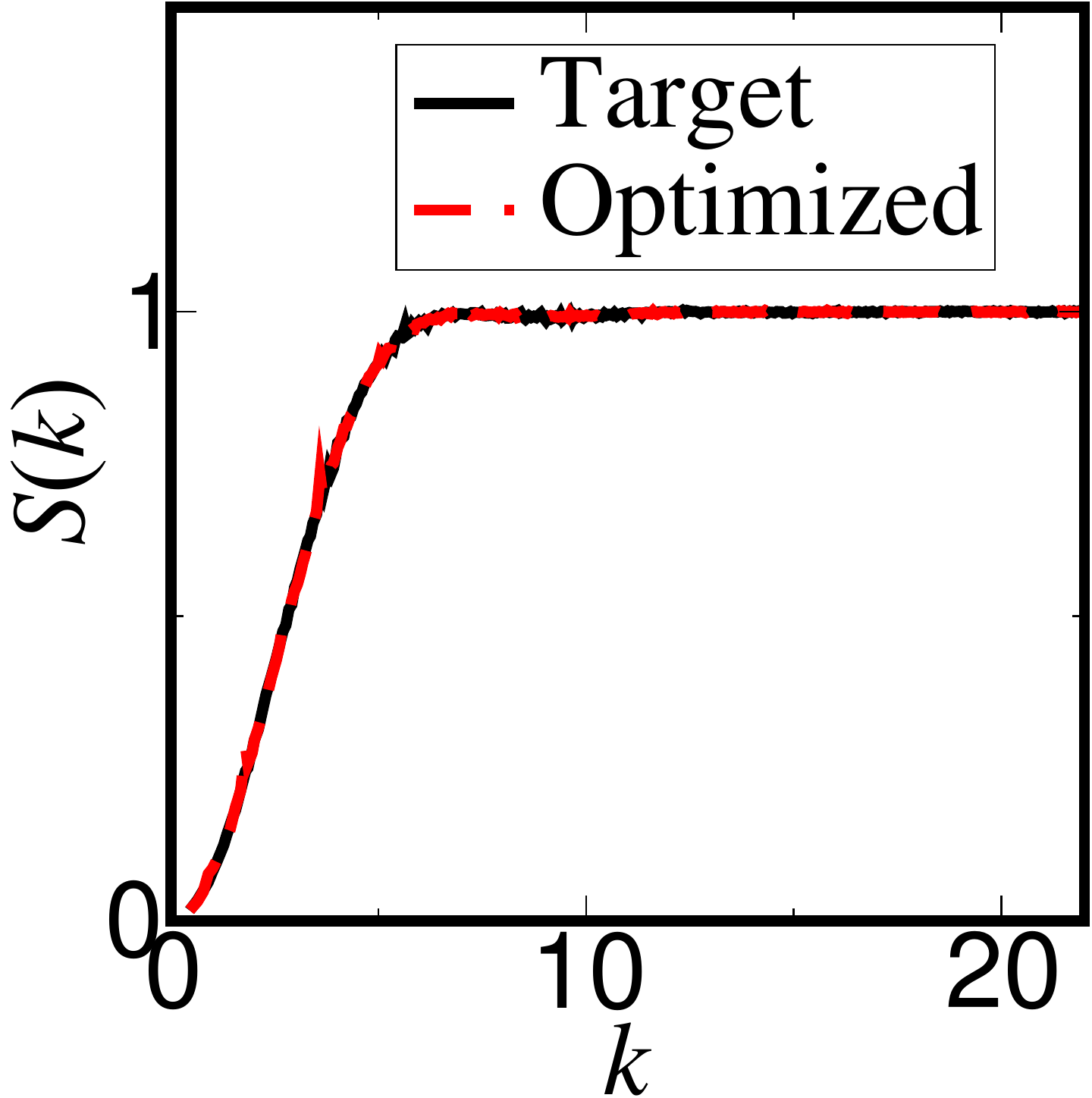}}
  \caption{(a) A portion of a 3D configuration of a 2,744-particle system that is equilibrated under the optimized effective one- and two-body potential for the target 3D cloaked URL. Only 512 particles are displayed. (b) Optimized pair potential [Eq. (\ref{url_v})] minus its long-ranged repulsive part $0.940/r$. (c) Targeted and optimized pair correlation functions. (d) Targeted and optimized structure factors.}
  \label{3DURL_basic}
\end{figure*}

\section{Conclusions and Discussion}
\label{conclusions}


We have formulated a novel optimization algorithm  to find effective one- and two-body potentials with high precision
that correspond to pair statistics
of general translationally invariant disordered many-body equilibrium or nonequilibrium systems  at positive temperatures.
The versatility and power of our inverse methodology to accurately extract the effective pair interactions
was demonstrated by considering four diverse target systems: (1) 3D liquid under the Dzugutov potential; (2)
2D Lennard-Jones system in the vicinity of its critical point; 
(3) 2D nonequilibrium RSA packing; and 
(4) a 3D nonequilibrium hyperuniform ``cloaked" URL.
We showed that the optimized pair potentials generate corresponding pair statistics
that accurately match their corresponding targets with total $L_2$-norm errors that are an order of magnitude smaller than that of previous methods. 
The results of our investigation lend further support to the Zhang-Torquato conjecture \cite{Zh20}, which
states that any realizable  $g_2({\bf r})$ or $S({\bf k})$ corresponding to a translationally invariant nonequilibrium system can be attained by a classical equilibrium ensemble involving only (up to) effective pair interactions. 

The capability of our procedure to precisely determine equilibrium systems corresponding to nonequilibrium pair statistics has important implications. First, our procedure provides an effective means to test the Zhang-Torquato conjecture for structures that span diverse hyperuniform and nonhyperuniform classes \cite{Zh20}. Second, the dynamics leading to a nonequilibrium system that has the same pair statistics as one drawn from an equilibrium ensemble must be reflected  in differences in their respective higher-order statistics. Thus, such differences in the higher-order statistics is expected to provide a measure of the degree to which a nonequilibrium system
is out of equilibrium. Third, such investigations  will enable one to probe systems
with identical pair statistics but different higher-body
statistics, which is expected to  shed light on the well-known degeneracy problem of statistical mechanics \cite{Ge94, Ji10a,St19}. Fourth, one could explore thermodynamic and dynamic properties of such effectively equivalent equilibrium systems, such as phase behaviors, excess entropies \cite{Ha86, Ni21b}, and inherent structures \cite{St82}, which are  outstanding problems for future research. Fifth, structural properties of the effectively equivalent equilibrium states, such as nearest-neighbor probability distribution functions, percolation threshold and fluid permeability, enable one to infer these nontrivial attributes of the nonequilibrium states, which are crucial in determining mechanical and electronic properties of materials \cite{To02a}. Finally, our study enables one to generate tunable nonhyperuniform and hyperuniform materials.

Another promising application of our methodology is to numerically investigate the realizability problem, i.e., whether hypothetical functional forms of $g_2(r)$ or $S(k)$ at fixed $\rho$ can be attained by many-particle configurations \cite{Ya61,Cos04,Uc06a,To06b,Ku07,Zh20}. This is an outstanding problem in statistical physics, as it has been shown that certain functional forms of pair statistics are not realizable, even if they meet all the explicitly known realizability conditions \cite{Ya61,To06b}. Our procedure provides a powerful means to test whether there exist equilibrium many-body systems with up to pair interactions that realize hypothetical target pair statistics at all length scales.

\begin{figure}[htp]
  \centering
  \includegraphics[width=55mm]{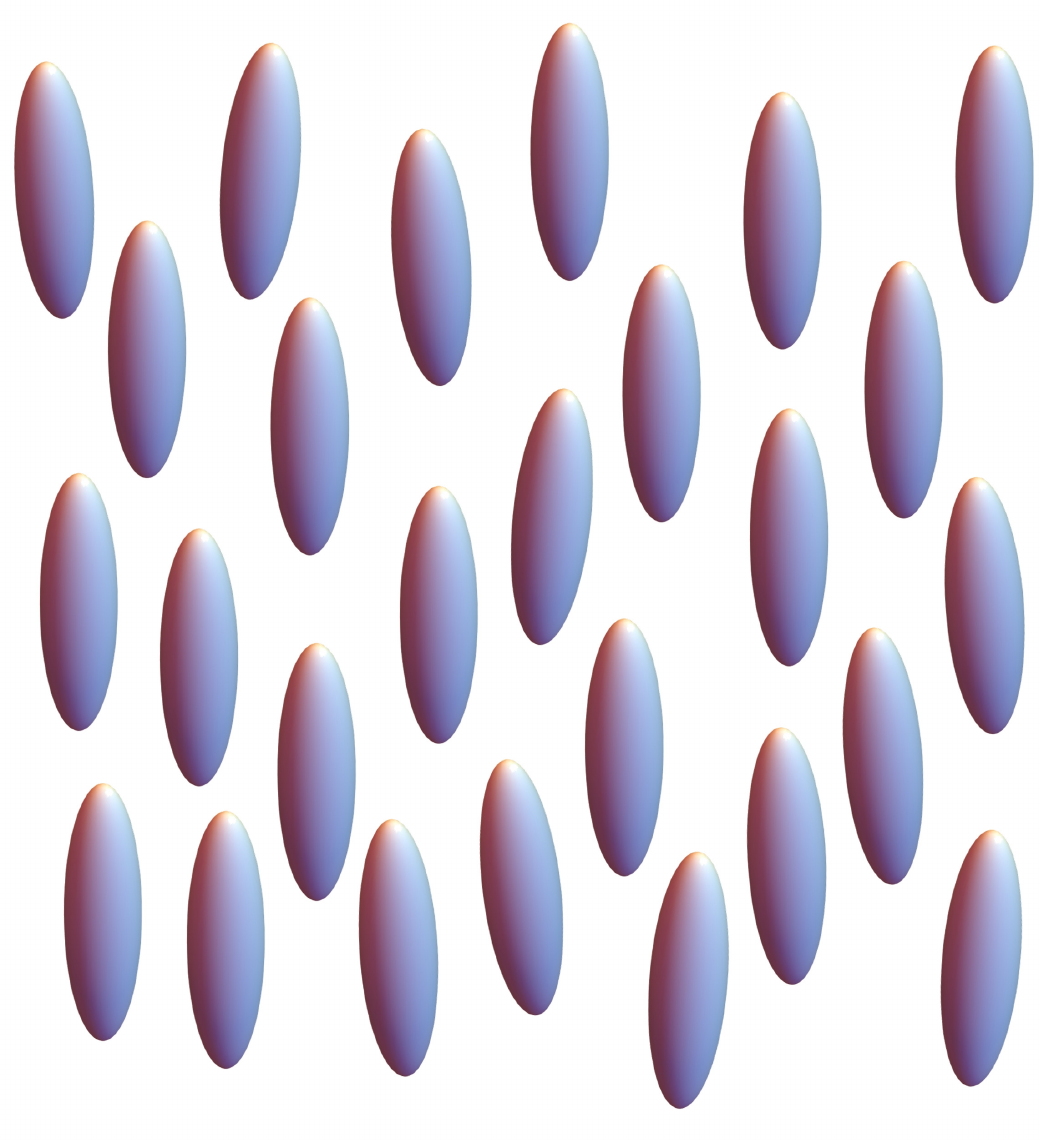}
  \caption{Schematic illustration of a statistically homogeneous but anisotropic nematic liquid-crystal configuration.}
  \label{nematic}
\end{figure}

The generalization of our methodology to anisotropic systems described in Sec. \ref{aniso} can be applied to determine effective one-and two-body potentials that
reproduce target pair statistics for liquid crystal phases. Such a concrete example includes the nematic phase of oriented ellipsoids  shown in Fig. \ref{nematic}. In this particular case,
the shape of the ellipsoids can be inferred from the range of repulsion of the anisotropic effective pair potential $v(\mathbf{r};\mathbf{a})$. It is also
straightforward to extend our methodology to inverse problems for statistically 
homogeneous multicomponent systems that consist of different species $A, B, ...$. Here, the goal is to find equilibrium systems with 
up to pair potentials that reproduce all the intraspecies and interspecies pair correlation functions $\{g_{AA}(\mathbf{r}),
g_{AB}(\mathbf{r}), ..., g_{BB}(\mathbf{r}), ...\}$ as well as their corresponding structure factors. To accomplish this task, 
one must choose a different set of basis functions for each
species-specific trial pair potential $\{v_{AA}(\mathbf{r},\mathbf{a}_{AA}), v_{AB}(\mathbf{r},\mathbf{a}_{AB}), ..., v_{BB}(\mathbf{r};\mathbf{a}_{BB}),...\}$ based on statistical-mechanical theory
\cite{He76,Sc03b}. The
algorithm described in Sec. \ref{opt} can be then applied to optimize each species-specific pair potential.

A substantially more challenging extension of our methodology is to targeted pair statistics derived from 
general statistically inhomogeneous system in which there is a preferred origin in the system, e.g. liquid-gas interfaces, which requires a
position-dependent one-body potential $\phi(\mathbf{r})$ and a pair potential that depends on absolute positions $\phi_2({\bf r}_1,{\bf r}_2)$.
To treat such systems, one
could apply the inhomogeneous Ornstein-Zernike equations \cite{Ha86} and the associated closures, such as the 
Lovett–Mou–Buff–Wertheim equation  \cite{Lo76}, to obtain an initial
guess for the functional forms of both $\phi(\mathbf{r})$ and $\phi_2(\mathbf{r}_1,\mathbf{r}_2)$  \cite{Br08}. 
However, it will be significantly more nontrivial to find appropriate pointwise
functional forms for the basis functions and to perform the corresponding optimizations
for inhomogeneous systems, and so represents an outstanding subject for future research.

We note that machine-learning approaches \cite{Mo19} could be used as an alternative to the BFGS technique for optimization of the potential parameters $\mathbf{a}$ to improve computational efficiency. More specifically,  the training phase of machine-learning techniques, i.e., generating a sufficient number of $[v(r;\mathbf{a}), g_2(r;\mathbf{a}) \text{ and } S(k;\mathbf{a})]$ pairs for many different parameter supervectors $\mathbf{a}$, can be completely parallelized. Combining machine-learning techniques with our inverse algorithm is an outstanding problem for future research. 

Finally, we remark that while the main focus of the present work concerns classical systems, one could also consider inverse problems for targeted pair statistics of quantum many-body systems, such as those involving boson or fermion statistics \cite{Ch87a, Gi19} or superfluidity as exhibited by liquid helium at low temperature \cite{La41, Fe56}. One could start with some experimentally measured $g_2(\mathbf{r})$ or $S(\mathbf{k})$ for a quantum fluid and determine if there exist classical many-particle systems interacting with up to pair potentials that produce the same pair statistics, which is often the case. For instance, it is known that the pair statistics of the 1D free-fermion system are equivalent to those of a classical system on a unit circle under a logarithmic interaction at positive $T$, known as the circular unitary ensemble \cite{Su70, To08b}. Another possible problem is to determine (at constant $\rho$ and $T$) how pair potentials in an initially classical fluid must be modified to keep the same pair distribution as the particles reduce their mass to become increasingly quantized, either as fermions or bosons. Investigating inverse problems for quantum systems is a fascinating area for future research.

\begin{acknowledgements}
We are grateful to Oliver Philcox for valuable discussions.
This work was supported by the National Science Foundation
under Award No. CBET-1701843. S.T. thanks the Institute for Advanced Study for their
hospitality during his sabbatical leave there.
\end{acknowledgements}

\appendix*

\section{Implementation details of the methodology for each of the targets}

\label{details}

\subsection{Lennard-Jones fluid in the vicinity of the critical point}
\label{app_crit}
To estimate the large-$r$ behavior of $v(r)$, we first fitted the the direct correlation function of the target system $c_T(r)/\beta$ in the range $1.5 \leq r \leq 5$ with Eq. (\ref{gamma_form})--(\ref{power_law}), in which the oscillatory factors were eliminated as $c_T(r)$ is not oscillatory about the horizontal axis. The form with the lowest BIC was given by a power law $-c_T(r)/\beta \sim -E r^{-p}$, where $p=5.9\pm 1.2$. To more accurately estimate the large-$r$ behavior of $v(r)$, $\tilde{c}_T(k)$ in the range $0\leq k\leq 4$ was fitted with the expected small-$k$ functional forms of $\tilde{c}(k)$ associated with $p=5, 6, 7$, respectively. (We fitted $\tilde{c}_T(k)$ instead of $S_T(k)$, as the former gave lower fit residuals.) Table \ref{crit_tildeC} shows the results of the fits along with their corresponding BIC. Although the fit for $p=5$ yielded the lowest BIC, its corresponding nonanalytic term in $\tilde{c}(k)$ is positive, which is inconsistent with the desired attractive large-$r$ behavior of $v(r)$. Thus, the large-$r$ asymptotic behavior of $v(r)$ is best described by $v(r)\sim -E/r^6$, where $E$ is estimated to be 3.88 via a numeric fit using Eq. (\ref{s_smallk}), but with $k^\zeta$ replaced by $\ln(k)k^\zeta$. To choose a functional form for the small-$r$ behavior of $v(r)$, we observed that the HNC approximation in the range $0.9 \leq r \leq 1$ is strongly repulsive and can be best described by a power law function (\ref{power_law}). Since $v_{\text{HNC}}(r)$ is not oscillatory about the horizontal axis, we eliminated the oscillatory factor $\cos\left(r/\sigma_j^{(1)} + \theta_j\right)$. We found that $v_{\text{HNC}}(r)$ in the range $0.9<r<10$ can be well fitted by the parameterized trial potential Eq. (\ref{ljGeneral}), in which the initial values of the parameters are $\varepsilon_1=3.90, p_1=11.51, \varepsilon_2=3.88$.

\begin{center}
\begin{table*}
\caption{Fits of $\tilde{c}_T(k)$ for the target near-critical LJ fluid in the range $0\leq k \leq 4$.}
\begin{tabular}{ ||c|c|c|c|c|c|c|c|| } 
 \hline
 Form of $v(r)$ at large $r$ & Form of $\tilde{c}(k)$ at small $k$ & $m_0$ & $m_2$ & $m_4$& $m_6$ & $m^*$ & BIC \\
 \hline
 $-r^{-5}$ & $\sum_{i=0}^3 m_{2i} k^{2i} + m^* k^3$ & 2.30& -5.64 & -0.292 & $9.76\times 10^{-4}$ & 2.44 & -252 \\
 $-r^{-6}$ & $\sum_{i=0}^3 m_{2i} k^{2i} + m^* \ln(k) k^4$ & 2.21& -4.44 & 1.08 & 0.00862 & -0.692 & -248 \\
 $-r^{-7}$ & $\sum_{i=0}^3 m_{2i} k^{2i} + m^* k^5$ &  2.14 & -4.02 & 1.08 & 0.0307 & -0.333& -188\\
 \hline
\end{tabular}
\label{crit_tildeC}
\end{table*}
\end{center}

In the optimization stage of the inverse procedure, we set the parameters defined in Sec. \ref{meth} to be $\sigma_{g_2}=4, \sigma_{S}=2, A=1, \lambda=1, \delta a=0.1, \epsilon=0.002, \epsilon'=0.02$. The criterion $\Psi<\epsilon$ is achieved within one single stage of optimization. The optimized $v(r;\mathbf{a})$ gives $\varepsilon_1=3.98, p_1=11.93, \varepsilon_2=4.00$.

\subsection{Liquid under the Dzugutov potential}
The target-generating Dzugutov potential is given by \ref{Dzu_def}, where the parameters used in this work are listed in Table \ref{dzu_targ_params}. To apply our methodology, we first note that $S_T(k)$ in the range $0\leq k\leq 1.5$ is fitted accurately by $0.169+0.0251k^2+0k^4$, which suggests $v(r)$ at large $r$ has exponential or superexponential decay. We accurately obtained the small-$k$ behavior for $\tilde{c}_T(k)$ and performed inverse Fourier transform (\ref{inverse_fourier_radial}), from which we determined that $-c_T(r)/\beta \sim v(r)$ has an effectively superexponential decay at large $r$.

\begin{center}
\begin{table}
\caption{Parameters of the target-generating Dzugutov potential [Eq.(\ref{Dzu_def})].}
\begin{tabular}{ ||c|c||c|c|| } 
 \hline
 $A$ & 5.92 & $b$ & 1.94 \\
 $B$ & 1.28 & $c$ & 1.1 \\
 $m$ & 16 & $d$ & 0.27 \\
 $a$ & 1.87 &  & \\
 \hline
\end{tabular}
\label{dzu_targ_params}
\end{table}
\end{center}

To obtain an initial guess of $v(r;\mathbf{a})$ for small- and intermediate-$r$, we observe that $v_{\text{HNC}}(r)$ has a strong repulsion for $r<1$ and is oscillatory in the range $1<r<2$. We fitted $v_{\text{HNC}}(r)$ in the range $0\leq r \leq 5$ with the forms (\ref{gamma_form})--(\ref{power_law_pot}) and found the fit with the lowest BIC is given by the exponential-damped oscillatory form \ref{exponential_form}. Thus, we started with an initial $v_(r;\mathbf{a})$ that is a sum of a power-law function and two exponential-damped basis functions" with $M=2$ and 4, respectively. This combination of $M$ values  achieved the lowest BIC among all combinations.
\begin{widetext}
\begin{equation}
    v(r;\mathbf{a})=
    \frac{\varepsilon_1}{r^{p_1}} 
    + \varepsilon_2\exp\left[-\left(\frac{r}{\sigma_2{(1)}}\right)^2\right]\cos\left(\frac{r}{\sigma_2^{(2)}}+\theta_2\right) + \varepsilon_3\exp\left[-\left(\frac{r}{\sigma_3^{(1)}}\right)^4\right]\cos\left(\frac{r}{\sigma_3^{(2)}}+\theta_3\right)
\label{dzu_v_ini}
\end{equation}
\end{widetext}

In the optimization stage of the inverse procedure, we used the parameters $\sigma_{g_2}=4, \sigma_{S}=2, A=1, \lambda=0.05, \delta a=0.1, \epsilon=0.002, \epsilon' = 0.02$. The optimization using the initial set of basis functions (\ref{dzu_v_ini}) stalled at $\Psi=0.011$, which is larger than the tolerance $\epsilon$. Thus, we added basis functions via a fit of the difference between target and trial potentials of main force (\ref{pmf}). The additional basis functions include another power-law function to accurately capture the repulsive interactions, as well as three undamped oscillatory functions in the range $0\leq r \leq 2$ to capture the asymmetric peaks and valleys in $g_{2.T}(r)$ in that range. Upon re-optimization of the parameters, the potential form (\ref{dzu_v_opt}) achieved the desired error tolerance $\Psi<0.002$. The optimized parameters are listed in Table \ref{dzu_opt_params}.

\begin{center}
\begin{table}
\caption{Optimized parameters of the pair potential (\ref{dzu_v_opt}) for the 3D Dzugutov liquid.}
\begin{tabular}{ ||c|c||c|c|| } 
 \hline
 $\varepsilon_1$ & 3.743 & $\theta_4$ & 0.7713\\ 
 $p_1$ & 11.108 & $\varepsilon_5$ & -1.072\\ 
 $\varepsilon_2$ & 0.7342 & $\sigma_5$ & 0.1619\\
 $p_2$ & 16.641 & $\theta_5$ & 4.455\\
 $\varepsilon_3$ & 45.126 & $\varepsilon_6$ & 2.893\\
 $\sigma_3^{(1)}$ & 0.7738 & $\sigma_6$ & 0.2416\\
 $\sigma_3^{(2)}$ & 0.1227 & $\theta_6$ & 1.349\\
 $\theta_3$ & 0.1501 & $\varepsilon_7$ & 6.017\\
 $\varepsilon_4$ & 24.704 & $\sigma_7$ & 1.036\\
 $\sigma_4^{(1)}$ & 0.1096 & $\theta_7$ & 3.193\\
 $\sigma_4^{(2)}$ & 0.1686 & & \\
 \hline
\end{tabular}
\label{dzu_opt_params}
\end{table}
\end{center}

\subsection{Equilibrium system corresponding to nonequilibrium random sequential addition pair statistics}
\label{app_rsa}
To choose an initial set of basis functions, we first observe that the corresponding $\tilde{c}_T(k)$ in the range $0\leq k \leq 3$ can be excellently fitted by the analytic function $-24.4 +1.51 k^2 + 0.00668 k^4$. This implies that $v(r)$ at large $r$ has exponential or superexponential decay. Thus, we fitted $v_{\text{HNC}}(r)$ in the range $1<r<8$ with Eqs. (\ref{gamma_form})-- (\ref{Yukawa_form}), and found that the fit with lowest BIC was achieved by a sum of two oscillatory Gamma functions and one non-oscillatory Gamma function. Our initial choice of the parameterized potential is given by
\begin{equation}
   v(r;\mathbf{a})=
    \begin{cases}
    \infty \qquad &r\leq 1\\
    -\frac{\varepsilon_1}{\Gamma\left(r/\sigma_1\right)} + \frac{\sum_{j=2}^{3}\varepsilon_j\cos\left(r/\sigma_j^{(1)} + \theta_j\right)}{\Gamma
    \left(r/\sigma_2^{(2)}\right)} \qquad &r>1.
    \end{cases}
    \label{rsapot0}
\end{equation}

In the optimization stage of the inverse procedure, we used the parameters $\sigma_{g_2}=4, \sigma_{S}=2, A=1, \lambda=0.5, \delta a=0.1, \epsilon=0.002, \epsilon' = 0.02$. The optimization using the initial set of basis functions (\ref{rsapot0}) stalled at $\Psi=0.01$, which is larger than the tolerance $\epsilon$. We observed that this was due to two reasons: (a) the optimized $g_2(1^+;\mathbf{a})$ was lower than the target $g_{2,T}(1^+)$, and (b) the maximum in $g_{2,T}(r)$ at $r=2.2$ was not well reproduced: the optimized $g_2(r;\mathbf{a})$ had a maximum at $r=2.05$ instead. In order to fix these discrepancies, in the second stage, we included in $v(r;\mathbf{a})$ one more oscillatory and one more non-oscillatory Gamma functions, i.e.,
\begin{equation}
    v(r;\mathbf{a})=
    \begin{cases}
    \infty \qquad &r\leq 1\\
    -\sum_{j=1}^{2}\frac{\varepsilon_j}{\Gamma\left(r/\sigma_j\right)} + \frac{\sum_{m=3}^{5}\varepsilon_j\cos\left(r/\sigma_j^{(1)} + \theta_j\right)}{\Gamma
    \left(r/\sigma_3^{(2)}\right)} \qquad &r>1.
    \end{cases}
    \label{rsapot1}
\end{equation}
Re-optimizing the parameters in Eq. (\ref{rsapot1}) still did not reach the desired convergence criterion, but this time solely due to the reason (b) described above. Therefore, we iteratively added Gamma-damped oscillatory functions and optimized the parameters, until $\Psi < 0.002$ was achieved. The final form of the parameterized pair potential is given by Eq. (\ref{rsapot}) and the optimized parameters are listed in Table \ref{rsa_params}.

\begin{center}
\begin{table}
\caption{Optimized parameters of the RSA effective pair potential (\ref{rsapot}).}
\begin{tabular}{ ||c|c||c|c|| } 
 \hline
 $\varepsilon_1$ & $1.200\times 10^{62}$ & $\epsilon_5$ & 0.5765\\ 
 $\sigma_1$ & 0.0200 & $\sigma_5^{(1)}$ & 0.223\\ 
 $\varepsilon_2$ & 7.610 & $\theta_5$ & 0.5056\\
 $\sigma_2$ & 0.214 & $\varepsilon_6$ & -0.2232\\
 $\varepsilon_3$ & 0.05790 & $\sigma_6^{(1)}$ & 0.213\\
 $\sigma_3^{(1)}$ & 0.0912 & $\theta_6$ & 1.352\\
 $\theta_3$ & 1.918 & $\varepsilon_7$ & 0.5458\\
 $\varepsilon_4$ & 0.3947 & $\sigma_7^{(1)}$ & 0.292\\
 $\sigma_4^{(1)}$ & 0.190 & $\theta_7$ & 4.395\\
 $\theta_4$ & 1.329 & $\sigma_3^{(2)}$ & 0.697\\
 \hline
\end{tabular}
\label{rsa_params}
\end{table}
\end{center}

\newpage
\subsection{Equilibrium system corresponding to nonequilibrium 3D cloaked URL system}
\label{app_url}
The fact that the target is hyperuniform with $\alpha=2$ implies that the pair potential has the asymptotic form $v(r)\sim 1/r$ [Eq. (\ref{v_hu})]. To obtain an initial form of the small- and intermediate-$r$ behaviors, we fitted $v_{\text{HNC}}(r)$ in the range $0\leq r \leq 2$ with Eqs. (\ref{gamma_form})--(\ref{power_law}), where the oscillatory factors were eliminated as $v_{\text{HNC}}(r)$ is not oscillatory about the horizontal axis. The form with lowest BIC was achieved by Eq. (\ref{url_v}). 

During the optimization step, we used the parameters $\sigma_{g_2}=4, \sigma_{S}=2, A=1, \lambda=2, \delta a=0.05, \epsilon=0.002$ and $\epsilon'=0.02$. The criterion $\Psi<\epsilon$ was achieved in one single stage of optimization. The optimized parameters are listed in Table \ref{URL_params}.

\begin{center}
\begin{table}
\caption{Optimized parameters of the 3D cloaked URL effective pair potential (\ref{url_v}).}
\begin{tabular}{ ||c|c||c|c|| } 
 \hline
 $\varepsilon_1$ & 0.940 & $\varepsilon_3$ & 0.0910 \\
 $\varepsilon_2$ & 1.57 & $\sigma_3^{(1)}$ & 0.650 \\
 $\sigma_2$ & 0.195 & $\sigma_3^{(2)}$ & 0.292 \\
 $p_2$ & 0.790 &  & \\
 \hline
\end{tabular}
\label{URL_params}
\end{table}
\end{center}

\FloatBarrier
%

\end{document}